\documentclass[10pt,twocolumn,prb,english,aps,showpacs,floatfix,groupedaddress,superscriptaddress]{revtex4-1}

\usepackage{graphicx}
\usepackage{epsfig}
\usepackage[english]{babel}
\usepackage{amsmath}
\usepackage{amssymb}
\usepackage{amsfonts}
\usepackage{longtable}
\usepackage{dcolumn}
\usepackage{bm}
\usepackage{bbm}
\usepackage{color,array}
\usepackage{colortbl}

\usepackage{bbold}
\usepackage{ctable}

\newcommand{\ket}[1]{\left|#1\right\rangle}		
\newcommand{\bra}[1]{\left\langle#1\right|}		

\begin{document}
\title{Multi-Particle Spectral Properties in the Transverse Field Ising Model by Continuous Unitary Transformations}

\author{Benedikt Fauseweh}
\email{benedikt.fauseweh@tu-dortmund.de}
\author{G\"otz S. Uhrig}
\email{goetz.uhrig@tu-dortmund.de}
\affiliation{Lehrstuhl f\"{u}r Theoretische Physik I, TU Dortmund, Otto-Hahn Stra\ss{}e 4, 44221 Dortmund, Germany}

\date{\rm\today}

\begin{abstract}
The one-dimensional transverse field Ising model is solved by continuous unitary
transformations in the high-field limit. A high accuracy is reached due to
the closure of the relevant algebra of operators which we call string operators.
The closure is related to the possibility to map the model by Jordan-Wigner transformation to non-interacting
fermions.  But it is proven without referring to this mapping.
The effective model derived by the continuous unitary transformations
is used to compute the contributions of one, two, and three elementary excitations
to the diagonal dynamic structure factors. The three-particle contributions have so far not
been addressed analytically, except close to the quantum critical point.
\end{abstract}




\maketitle

\section{Introduction} 
\label{sec:intro}

{Understanding strong quantum fluctuations continues to be a formidable challenge.
Where two (or more) phases compete and are separated by a continuous
quantum phase transition \cite{Sachdev2}, i.e.,  at zero temperature, such 
fluctuations are particularly strong. Generically, such a quantum phase transition 
is signaled by the decay of elementary excitations. Far away from the
phase transition, spectroscopic probes show dominant sharp $\delta$-peaks at low energies which
result from stable elementary excitations, so-called quasi-particles.
But on approaching the phase transition, the spectral weight in the dominant
quasi-particle peak is reduced further and further and shifted to contributions
of more quasi-particle. Multi-particle spectra with considerable weight
are an important signature of dominant quantum fluctuations in general.
The vanishing of the single quasi-particle peaks at zero temperature is a smoking gun
for a quantum phase transition in particular.
}

In this context, the transverse field Ising model (TFIM){\cite{katsu62}} is a popular 
{generic} model
describing magnetic excitations and displaying a quantum phase transition
between the disordered phase in the limit of dominating transverse field and
the ordered phase in the limit of dominating longitudinal Ising coupling.
Due to its relative simplicity, the TFIM provides a {convenient} test case in the 
development of theoretical approaches \cite{vidal07,gCUT}.
This is particularly true for the one dimensional case, for which fermionization
by a Jordan-Wigner transformation \cite{JordanWignerTrafo} yields an
exact solution \cite{Niemeijer,Pfeuty,DSF2}. 

{The calculation of dynamical correlations in the TFIM is an active  field of research. 
While transverse correlations can be treated in terms of fermions \cite{DSF2,DSF}, longitudinal correlations require different approaches, because of their non-locality in the fermionic picture. 
Based on an equation of motion for the longitudinal correlations \cite{perk80} there has
been a series of papers investigating the scaling region around the critical point
Ref.\ \onlinecite{mccoy83a,mccoy83b,mulle83,mulle84a,mulle84b,mulle85}. In 2009, 
Perk and Au-Yang computed results for the time-dependent longitudinal correlation functions
by solving the coupled differential equations and complementing them with long-time asymptotics \cite{perk09}.
But so far no momentum and frequency resolved analysis has been performed, which also applies
away from the scaling regime.}

Quantum magnets in the vicinity of quantum phase transitions 
are dominated by strong quantum fluctuations. Quantum fluctuations are strongly
favored by low-dimensionality and by frustration. Theoretically, a particularly
clear sign of dominant quantum fluctuations is the fractionalization of elementary
excitations. For instance, conventional spin waves (magnons) split into two
spinons in antiferromagnetic Heisenberg chains \cite{fadde81,nagle91,dende96,karba97}.
Before complete fractionalization {occurs}, the spectral weight, as observed in
inelastic neutron scattering, shifts from the channel of a single elementary excitation
to the channel where two and more elementary excitations are created \cite{knett01b,schmi05b}.
Thus, also the experimental focus is directed more and more to continua formed by more 
than one excitation, see for instance Refs.\ \onlinecite{nagle91, dende96, notbo07, chris07}.

In view of the above considerations, the present article pursues two goals
in a study of the one-dimensional (1D) TFIM.
First, we show how the special algebraic structure (`string algebra') 
of the operators occurring in the 1D TFIM enables its solution 
by a continuous unitary transformation (CUT) in the high-field phase to very high 
accuracy. {Upon completion of our calculations, we learned that this algebra was introduced
and used before in an algebraic solution of the TFIM \cite{jha73}.}

This {algebra} paves the way to treat a larger class of models of which 
the Hamilton operators can be expressed by operators belonging to the 
string algebra, for instance  $XY$ models in transverse fields.
Second, we compute the three-particle contributions to the diagonal dynamic
structure factors (DSF) in the CUT framework in the high-field phase.
To our knowledge, this is the first time that these subdominant
contributions are computed, except in the scaling
region around the quantum phase transition. 
Thereby, interesting predictions for future experimental studies are provided.

In Sect.\ II, the model and known exact results are recalled. In the following 
section, the continuous unitary transformations (CUTs) are briefly reviewed.
Sect.\ IV is devoted to the algebra of string operators which are subsequently employed.
Sects.\ V and VI comprise our static and dynamic results, respectively, while the
conclusions are drawn in Sect.\ VII.

\section{Model and exact results} 
\label{sec:model}

The Hamiltonian for the transverse field Ising model (TFIM) reads
\begin{align}
H_\mathrm{TFIM} 
&= \frac{\Gamma}{2} \sum\limits_i \sigma_i^z + \frac{J}{4} \sum\limits_i \sigma_i^x \sigma_{i+1}^x , \label{eq:hamiltonian}
\end{align}
where the $\sigma^\alpha$ are the Pauli matrices and the sum $i$ runs over all lattice sites. We normalized the distance between two sites to one. The transverse field strength is given by the parameter $\Gamma$ while $J$ denotes the strength of the antiferromagnetic coupling between two adjacent sites. The antiferromagnetic exchange can be {converted to} a ferromagnetic exchange 
$J \rightarrow -J$ by a $\pi$ rotation around $S_i^z$ for every second site $i$. This translates to a shift of $\pi$ in  momentum space.

The model has a quantum phase transition at $J=2\Gamma$ and it is self dual \cite{Sachdev2}. Similar to Ref.\ \onlinecite{Sachdev2}, we introduce the parameter $x = J/2\Gamma$. The starting point for the {CUT} calculations is $J=0$. Hence, an elementary excitation is given by a single spin flip. 
{For finite} $J$ the energy of these excitations becomes {momentum dependent}. 
We will refer to these excitations as \emph{quasi-particles}. We expect that the perturbative ansatz breaks down once we reach the critical value $J=2\Gamma$. Hence we focus on the static and dynamic properties for $J<2\Gamma$.

The model was solved exactly by Pfeuty in 1970 \cite{Pfeuty}, based on the work by Lieb et al. \cite{Lieb} and Niemeijer \cite{Niemeijer}. Pfeuty's solution uses the Jordan-Wigner 
transformation \cite{JordanWignerTrafo}
to map the Hamiltonian in Eq.\ \eqref{eq:hamiltonian} to a chain of free fermions 
which is diagonalized by  a Bogoliubov transformation \cite{Bogo}.
This approach yields the exact expression for the ground state energy per site
\begin{align}
\frac{E_0}{N \Gamma} = - \frac{1}{2 \pi} \int\limits_0^\pi \omega(q) \mathrm{d}q  ,
\end{align}
where $\omega(q)$ denotes the dimensionless dispersion
\begin{align}
\omega(q) = \sqrt{1 + x^2 - 2 x \cos(q) } .
\end{align}
The {dispersion with dimension} is given by $\Gamma \omega(q)$.

From the dispersion we can easily extract the energy gap of the lowest lying excitations
\begin{align} 
\label{eq:gap_exact}
\frac{\Delta}{\Gamma} = \left| 1 - x \right| .
\end{align}
Another interesting quantity worked out by Pfeuty is the transverse magnetization
\begin{align}
M_z  = \frac{1}{N} \sum\limits_i \bra{g} \sigma_i^z \ket{g} = \frac{1}{\pi} \int\limits_0^\pi \frac{1+x\cos(q)}{\omega(q)} \mathrm{d}q,
\end{align}
where $\ket{g}$ denotes the ground state of the TFIM.
In the following sections we will compare our results with these exact expressions in order to validate the 
CUT approach. 

Beside the static properties stated above, dynamic properties are important 
in order to explain experimental results. 
Although the TFIM is analytically integrable, the evaluation of longitudinal dynamic correlations remains a very difficult task which requires {considerable} numerics, see for instance Refs.\ \onlinecite{perk09} or \onlinecite{StolzeKrones}. 
{In the fermionic picture, this is due to the non-locality of the Jordan-Wigner transformation}.

One important {quantity} in the study of spin systems is the 
{dynamic structure factor (DSF)}
\begin{align} 
\label{eq.DSF_definition}
S^{\alpha \beta}(\omega, Q) &= \frac{1}{N} \int_{-\infty}^{\infty} \frac{\mathrm{d} t}{2 \pi} \sum\limits_{l, l'} e^{i \omega t} e^{-i Q(l-l')} \langle \sigma_l^\alpha(t) \sigma_{l'}^\beta \rangle ,
\end{align}
where $\alpha , \beta \in \left\lbrace x,y,z \right\rbrace$. {Here $Q$ denotes the total momentum and $\omega$ the frequency.} The DSF is directly linked to the differential cross section in inelastic scattering experiments, see for instance Ref.\ \onlinecite{Lovesey}. 

Due to the symmetry $\sigma_i^x \rightarrow -\sigma_i^x$ of $H_\mathrm{TFIM}$ 
no correlations occur for $\alpha=x, \beta=z$ and $\alpha=y, \beta=z$ and vice versa, but for $\alpha=x, \beta =y$ and for $\alpha=\beta$ the DSF may and will obtain finite values.

In the following, we focus on the diagonal part of the DSF, i.e., $\alpha = \beta$.
For $\alpha=z$ exact expressions are known \cite{DSF3, DSF2, DSF}, because the observable $\sigma^z$ {remains} 
local in the Jordan-Wigner representation of the TFIM. At zero temperature case, this  expression reads
 \begin{subequations}
\begin{align} 
\label{Eq.DSF_zz_exact}
S^{zz}(Q,\omega) &= \int\limits_{-\pi}^{\pi} \mathrm{d} k_1 \left[ 1-f(Q,k_1)  \right]  \nonumber \\
 				&\delta( \omega - \omega(k_1-Q/2) - \omega(k_1+Q/2) ) ,
\end{align}
with
\begin{align}  
f(Q,k_1) &= \frac{\left(\Gamma + \frac{J}{2}\cos(k_1-Q/2)\right) \left(\Gamma + \frac{J}{2}\cos(k_1+Q/2)\right) }{\omega(k_1-Q/2) \omega(k_1+Q/2)} .
\end{align}
\end{subequations}
It consists of a spectral density of scattering states of two elementary excitations with 
{total} momentum $Q$.
For $\alpha=x$ and $\alpha=y$, only the one-particle contributions have been calculated by Hamer et al.\ in 2006 \cite{HamerTFIM}. They used series expansion techniques to propose the expressions
\begin{subequations}
\begin{align}
\label{S1xx}
S_1^{xx}(Q) &= \frac{\left[1-x^2 \right]^{\frac{1}{4}} }{\omega(Q)} , \\
\label{S1yy}
S_1^{yy}(Q) &= \left[1-x^2 \right]^{\frac{1}{4}} \omega(Q) 
\end{align}
\end{subequations}
for the {one-particle contribution to the} equal-time structure factor. 
By comparing their results to correlation functions in the two-dimensional classical Ising model, 
see Ref.\ \onlinecite{2dIsing1, 2dIsing2}, they could show that the expressions above are indeed 
exact. Hence the full one-particle structure factor is given by
\begin{align}
\label{eq:equal_time_dsf}
S_1^{\alpha \alpha}(Q, \omega) = S_1^{\alpha \alpha}(Q) \delta(\omega - \omega(Q)) .
\end{align}

For higher-particle contributions to the longitudinal DSF much less is known. In 1978, Vaidya and Tracy computed exact expressions for the longitudinal correlation functions in the anisotropic $XY$ model in the time domain
\cite{Vaidya19781}. They evaluated the resulting expression in frequency space up to the three-particle contributions. But their results are limited to the scaling region at low energies, very close to the critical point. 
{Furthermore M\"uller and Shrock calculated frequency-integrated wave
number dependent susceptibilities  for the TFIM at the
critical point in Refs.\ \onlinecite{mulle84b,mulle85}.}
Our results will be {complementary} to theirs.

\section{Continuous unitary transformations} 
\label{sec:cut}

We use the method of \emph{continuous unitary transformations} (CUT) to derive effective models 
which allow for an easier evaluation of {static} 
ground state properties and dynamic correlation functions. 
The idea of CUT was first introduced by Wegner \cite{Wegner} and independently by G\l{}azek and 
Wilson \cite{GlazekWilson1, GlazekWilson2}. 

The concept of CUT is to systematically finda unitary transformation that maps the Hamiltonian to a
diagonal representation. One introduces a family of unitary transformations
$U(l)$ depending {differentiably} on a parameter $l \in \mathbb{R}^+$.
The unitary transformation is characterized by its anti-Hermitian generator 
$\eta(l) = (\partial_l U(l)) U^\dagger(l) {=-\eta^\dag(l)}$. 
Then, a short calculation yields the \emph{flow equation}
\begin{align} 
\label{eq:FlowEquation}
\partial_l H(l) = \left[ \eta(l), H(l) \right]
\end{align}
for the $l$-dependent Hamiltonian $H(l)$. 
In general, it represents a system of coupled differential equations for the prefactors of all operators appearing in $H(l)$. We refer to it as the \emph{differential equation system} (DES).
Without further {truncation}, the DES generically comprises an infinite number of variables.
In practice, various truncation schemes help to keep the DES finite.
For $l \rightarrow \infty$ the Hamiltonian acquires its final form and it is denoted as the 
effective Hamiltonian
\begin{align}
H_\mathrm{eff} = H(l) \big|_{l=\infty} = U(\infty) H U^\dagger(\infty).
\end{align}
The convergence for $\ell \to \infty$ is assumed; it cannot be {proven} generally
{for infinite dimensional quantum systems} because it depends on the specific form 
of the generator as well as on the {employed} truncation scheme.


Note that observables $O$ also need to be transformed to effective observables {by}
the same unitary transformation. This results in the flow equation for observables 
\begin{align} 
\label{eq:flow_obs}
\partial_l O(l) = \left[ \eta(l), O(l) \right]
\end{align}
which yields the effective observable $O_\mathrm{eff}$ for $l \rightarrow \infty$. 

The generator characterizes the CUT and the flow of the Hamiltonian.
Thus, the choice of the generator is an important issue and it still represents an active field of research,
cf.\ Refs.\ \onlinecite{Wegner,Mielke,KnetterUhrig,fisch10a,dresc11,gCUT}.
In this paper we use the \emph{(quasi-)particle conserving} (pc) generator, which was first proposed by Mielke \cite{Mielke} in the context of banded matrices and independently by Knetter and Uhrig \cite{KnetterUhrig} for
many-body problems. By `quasi-particle' we mean the elementary excitation.
The pc generator directly aims at these quasi-particles. The goal is to
eliminate terms that do not conserve the number $\widehat Q$ of quasi-particles
\begin{align}
\left[ H_\mathrm{eff}, \widehat Q \right] = 0.
\end{align}
The pc generator is given in matrix representation in the {eigenbasis} of $\widehat Q$ by
\begin{align}
\eta_{\mathrm{pc},ij}(l) = \mathrm{sgn}(q_i - q_j) h_{ij}(l),
\end{align}
where $q_i$ denotes the {eigenvalues} of the operator $\widehat Q$.

An equivalent description of the pc generator can be given by decomposing the Hamiltonian into parts that create, $H^+(l)$, conserve, $H^0(l)$, and annihilate, $H^-(l)$, quasi-particles. Then the Hamilonian
reads
\begin{align}
H(l) = H^+(l) + H^0(l) + H^-(l)
\end{align}
and the quasi-particle conserving generator
\begin{align} 
\label{eq:pc_gen_simple}
\eta_\mathrm{pc} = H^+(l) - H^-(l).
\end{align} 
The convergence of the flow induced by this generator is proven for finite-dimensional systems;
extensions to infinite systems are also available \cite{dusue04a}.
The pc generator preserves the blockband diagonal structure, i.e., 
the maximum number of particles created or annihilated does not change during the flow \cite{Mielke,KnetterUhrig,heidb02b}.

The CUT method consists of two basic steps.
The commutator in Eq.\ \eqref{eq:FlowEquation} needs to be calculated, followed by the integration of the resulting flow equation. 
The latter can easily be done with standard numerical integration algorithms or even analytically.

In general, commuting $H$ with $\eta$ creates new types of terms which were originally not part of the Hamiltonian.
For systems in the thermodynamic limit, all sorts of new terms may arise
connecting more and more sites over larger and larger distances. 
In a numerical calculation we cannot treat an infinite number of operators, 
hence we have to restrict ourselves to operators which are physically relevant. In this paper we use the previously introduced \emph{directly evaluated enhanced perturbative CUT} (deepCUT) \cite{epCUT}. The idea of deepCUT is to truncate operators and contributions to the DES according to their effects in 
powers of a small {expansion} parameter $x$. Roughly speaking, the order $n$ in $x$ is the truncation criterion.
More precisely, a certain contribution to the DES is kept if it affects the
targeted quantities (here: ground state energy and one-particle dispersion) 
in  order $m\le n$ in $x$. Details can be found in Ref.\ \onlinecite{epCUT}.

Thus we write our initial Hamiltonian in the form
\begin{align}
H = H_0 + xV
\end{align}
where $H_0$ describes the unperturbed Hamiltonian and $V$ represents a perturbation. 
We expand the operators in the basis $\left\lbrace A_i \right\rbrace$
{which is chosen such that the effective Hamiltonian can be computed exactly \cite{epCUT}
up to order $n$ in the parameter $x$.} 
Then the flowing Hamiltonian can be denoted as
\begin{align} 
\label{eq:flowing_H}
H(l) = \sum\limits_i h_i(l) A_i
\end{align}
where the prefactors $h_i(l)$ depend on the flow parameter $l$. 
For the generator we choose the same operator basis with the same prefactors
\begin{align}
\eta(l) = \sum\limits_i \eta_i(l) A_i = \sum\limits_i h_i(l) \eta \left[ A_i \right]
\end{align}
where $\eta \left[ \cdot \right]$ is a superoperator applying the generator scheme.
For the pc generator, $\eta \left[ A_i \right]= A_i$ if $A_i$ creates more quasiparticles
than it annihilates, $\eta \left[ A_i \right]= -A_i$ if $A_i$ annihilates more quasiparticles
than it creates, otherwise $\eta \left[ A_i \right]=0$, cf.\  Eq.\ \eqref{eq:pc_gen_simple}.

With this definitions, we obtain the flow equations 
\begin{align} \label{eq:flow_h_i}
 \partial_l h_i(l) = \sum\limits_{j,k} D_{ijk} h_j(l) h_k(l) .
\end{align}
We call the $D_{ijk} \in \mathbb{C}$ the \textit{contributions} to the DES.
They are obtained {in a perturbative calculation up to order $n$}
by calculating the commutator in Eq.\ \eqref{eq:FlowEquation} 
and expanding the results in the chosen operator basis. 
{Note that the numerically evaluated DES also comprises
powers in $x$ beyond the order $n$ \cite{epCUT}.}

\section{String operators} 
\label{sec:string}

In the previous section, we explained how continuous unitary transformations 
are applied in a general context.
{Here we specify the approach for the transverse field Ising model.}
For the TFIM in the high-field limit, we use the the state with all spins down 
\begin{align}
\ket{0} = \ket{ \cdots \downarrow_{j-1} \downarrow_j \downarrow_{j+1} \cdots } ,
\end{align}
as the reference state, i.e., as the vacuum of elementary excitations.
This corresponds to the strong field limit $\Gamma \rightarrow \infty$ in the TFIM, which is also the starting point for a perturbative approach in the parameter $x=J/(2\Gamma)$. 
An elementary excitations, i.e., a quasi-particle, is created by the spin-flip operator $\sigma_l^+$. 
We denote this state by
\begin{align} 
\label{eq:single_ex_state}
\ket{l} = \sigma_l^+ \ket{0}.
\end{align}
It is obvious that no two excitations can be present at the same site so that the
quasi-particles behave like hardcore bosons. But multi-particle states can straightforwardly
be created by flipping spins at different sites.

These ideas suggest a basis of operators consisting of monomials
made from the local operators 
\begin{align}
\label{setop-multi}
\left\lbrace \sigma_j^+, \sigma_j^-, \sigma_j^+ \sigma_j^-, \mathbb{1} \right\rbrace ,
\end{align}
where $\sigma_j^+$ stands for particle creation, $\sigma_j^-$ for particle annihilation, and
$\sigma_j^+ \sigma_j^-$ counts whether a particle is present at site $j$ ($\sigma_j^+ \sigma_j^-=1$)
or not ($\sigma_j^+ \sigma_j^-=0$). This approach is in line with the general structure
explained in Ref.\ \onlinecite{StructureOperators}; we call it henceforth the multi-particle representation.

The number of such monomials grows exponentially with the number of sites
which are non-trivially involved {because at any site a quasi-particle may 
be created or annihilated or simply counted. For each additional site occurring in the course 
of the iterated commutations, the number of operators to be tracked grows roughly by a factor of 4 (neglecting reductions due to symmetry effects).} 
This is a major drawback if one aims at higher
orders. Therefore, we introduce a simpler modified operator basis, which we call \emph{string algebra},
which is more appropriate for the TFIM{, see also Ref.\ \onlinecite{jha73}}. 
We stress that the possibility to introduce a string algebra is connected to the Jordan-Wigner representation of the Hamiltonian in terms of non-interacting fermions.

The string algebra consists of operators which are given by the following product of Pauli operators
\begin{subequations}
\label{eq:t_n}
\begin{align}
T_n^{\phi \epsilon} &:= \underset{j}{\sum} \sigma_j^\phi \left( \overset{j+n-1}{\underset{k=j+1}{\prod}} \sigma_k^z \right) \sigma_{j+n}^\epsilon \\
					  &= \underset{j}{\sum} \sigma_j^\phi \sigma_{j+1}^z \sigma_{j+2}^z \cdots \sigma_{j+n-1}^z \sigma_{j+n}^\epsilon , 
\end{align}
\end{subequations}
with $\{ \phi, \epsilon \} \in \{+,-\}$ and $n \in \mathbb{N}$.
Each string operator consists of a product of adjacent $\sigma_z$ operators,
 framed by spin flip creation- and/or annihilation operators.
 The $\sigma_z$ operators form the string between the pair of spin flip operators.
 We  refer to $n$ as the spatial \emph{range} of an operator. 
 In contrast to the local set of operators used in the
 multi-particle representation \eqref{setop-multi} the string algebra
 is more transparently expressed by the set
 \begin{align} 
\label{eq:string_basis}
\left\lbrace \sigma_j^+, \sigma_j^-, \sigma_j^z, \mathbb{1} \right\rbrace .
\end{align}
{The key point of the string algebra is that excitations or annihilations of quasi-particles
occur \emph{only} at the end points of the string. Thus, for given end points,
there are only four string operators to be considered. If excitations or annihilation
could occur anywhere along the string one would have exponential growth of the number
of operators.}
 
 
 In Eq.\ \eqref{eq:t_n}, we  defined the translationally invariant form of string operators. 
 When dealing with local observables, it is also useful to introduce local string operators
\begin{subequations}
\label{eq:o_n}
\begin{align}
O_{j,n}^{\phi \epsilon} &:= \sigma_j^\phi \left( \overset{j+n-1}{\underset{k=j+1}{\prod}} \sigma_k^z \right) \sigma_{j+n}^\epsilon \\
					  &=  \sigma_j^\phi \sigma_{j+1}^z \sigma_{j+2}^z \cdots \sigma_{j+n}^\epsilon ,
\end{align}
\end{subequations}
with $\{ \phi, \epsilon \} \in \{+,-\}$ and $n \in \mathbb{N}$ .
Note that a translationally invariant string operator is given by the sum of local string operators.

It is also useful to define a string operator of range $0$ consisting of a single $\sigma^z$ matrix.
\begin{subequations}
\begin{align} 
T_0 &:= \underset{j}{\sum} \sigma_j^z ,\\
O_{j,0} &:= \sigma_j^z. 
\label{eq:string_zero}
\end{align}
\end{subequations}
For $n=1$, the definitions \eqref{eq:t_n} and \eqref{eq:o_n} correspond to a normal hopping term or pair  creation or annihilation operator. These operators cannot be distinguished from operators
in the multi-particle representation \cite{StructureOperators}.

For $n>1$, the situation is different. For example, in the case $n=2$, 
$\phi=+$ and $\epsilon=-$ can be re-expressed in multi-particle representation by
\begin{subequations}
\begin{align}
T_2^{+-} &= \underset{j}{\sum} \sigma_j^+ \sigma_{j+1}^z \sigma_{j+2}^- \\ 
		 &= \underset{j}{\sum} \sigma_j^+ (2 \sigma_{j+1}^+\sigma_{j+1}^- - 1) \sigma_{j+2}^- \\
		 &= \underset{j}{\sum} \left(2 \sigma_j^+ \sigma_{j+1}^+\sigma_{j+1}^-\sigma_{j+2}^- - \sigma_j^+ \sigma_{j+2}^- \right),
\end{align}
\end{subequations}
which is the sum of a quartic interaction term and a hopping term,
 because we re-expressed $\sigma_{j+1}^z = 2 \sigma_{j+1}^+\sigma_{j+1}^- - 1$.
 
 This simple example illustrates the computational advantage of the string algebra. If we tracked all operators in multi-particle representation, a single string operator of range $n$ would require $2^{n-1}$ multi-particle operators{, clarifying the previous statement on the exponential growth of the number of such monomials.} Therefore, if a model can be diagonalized within the string algebra,
  it is highly advantageous to describe all operators in terms of string operators.

With the above definitions, the Hamiltonian of the transverse field Ising model 
{is} formulated in terms of string operators
\begin{subequations}
\begin{align} 
H_\mathrm{TFIM}  &=  \frac{\Gamma}{2} \underset{j}{\sum} \sigma_j^z + \frac{J}{4} \underset{j}{\sum}  \left( \sigma_j^+ \sigma_{j+1}^- + \sigma_j^+ \sigma_{j+1}^+ + \mathrm{h.c.} \right) \\
 				&= \frac{\Gamma}{2} T_0 + \frac{J}{4} \left( T_1^{+-} + T_1^{-+} + T_1^{++} + T_1^{--} \right). \label{eq:tfim_in_strings}
\end{align}
\end{subequations}

Next, we study the action of hopping terms on {single particle-states}
\begin{subequations}
\begin{align}
T_n^{-+} \ket{l} &= \underset{j}{\sum} \sigma_j^- \sigma_{j+1}^z \sigma_{j+2}^z \cdots \sigma_{j+n}^+ \ket{l} \\
				 &= \underset{j}{\sum} \delta_{l,j} \sigma_{j+1}^z \sigma_{j+2}^z \cdots \sigma_{j+n}^+ \ket{0} \\
				 &= (-1)^{n-1} \ket{l+n}.
\end{align}
\end{subequations}
In the second line we used the property $\sigma_j^- \ket{l} = \delta_{l,j} \ket{0}$ and we know that $\sigma_{j}^z \ket{0} = - \ket{0}$, which yields the final result. If there is only one quasi-particle in the system, the only difference to conventional hopping is the factor $(-1)^{n-1}$. 
For {sub}spaces with more quasi-particles, we have to take into account that there may be particles on the sites $l+1, l+2 \dots l+n-1$. 
They modify the exponent of $(-1)$ and can thus change the sign of the resulting state.

In order to apply the deepCUT to the TFIM, we have to calculate the contributions to the DES in Eq.\ \eqref{eq:flow_h_i}. Thus, we calculate the commutator between two operators of the Hamiltonian $H$ and the generator $\eta$. In Appendix \ref{app:a} we show that the string algebra is closed under such commutations.
This means that the commutator of two string operators can again be written as a linear combination of string operators. The closure of the string algebra allows us to set up the flow equation in very high {order} because the number of operators to be  tracked grows only linearly for the Hamiltonian. For local observables within the string algebra it grows quadratically which is still 
a moderate {growth}. This is the key observation of the present article.

Explicitly calculating  all distinct commutators of string operators, see Appendix \ref{app:a}, allows us to determine all contributions to the DES analytically. In this way, we calculate the flow equation up to infinite order in $x$. We parametrize the Hamiltonian
\begin{align}
\label{eq.tfim_in_strings}
H_\mathrm{TFIM}(l) &= t_0(l) T_0 + \sum\limits_{n=1}^\infty t_n^{+-}(l) \left( T_n^{+-} + \mathrm{h.c.} \right) \nonumber \\ &+ \sum\limits_{n=1}^\infty t_n^{++}(l) \left( T_n^{++} + \mathrm{h.c.} \right) ,
\end{align}
and the generator for the CUT
\begin{align}
\label{eq.generator_in_strings}
\eta(l) &= \sum\limits_{n=1}^\infty t_n^{++}(l) \left( T_n^{++} - \mathrm{h.c.} \right) .
\end{align}
In Appendix \ref{app:b} we derive the flow equation for the prefactors $t_0, t_n^{+-}, t_n^{++}$. It reads
\begin{subequations}
\label{eq:flow_infinite}
\begin{align}
\partial_l t_0^{\phantom{+-}} &= 2 \sum\limits_{n=1}^{\infty} \left(t_n^{++}\right)^2 ,\\
\partial_l t_m^{+-} &= 2 \sum\limits_{k,l = 1}^{k+l=m} t_k^{++} t_l^{++}  - 2 \sum\limits_{k,l = 1}^{|k-l|=m} t_k^{++} t_l^{++} , \\
\partial_l t_m^{++} &= -4 t_m^{++} t_0 + 2 \sum\limits_{k,l = 1}^{|k-l|=m} \mathrm{sgn}(k-l) t_k^{++} t_l^{+-} \nonumber \\ &+  2 \sum\limits_{k,l = 1}^{k+l=m} t_k^{++} t_l^{+-} ,
\end{align}
\end{subequations}
with $m \in \mathbb{N}$. Note that this is a differential equation with an infinite number of variables
which grows, however, only linearly in the spatial range. This result is remarkable considering the fact 
that it would require tremendously more flow parameters if we formulated the problem in multi-particle representation. Thus the string algebra allows us to evaluate the Hamiltonian transformation up to
very high orders, which is especially important {on approaching} the quantum {critical} point $x=J/2\Gamma=1$.

\section{Static results} 
\label{sec:static_results}

In this section we  evaluate and present static results for the transverse field Ising model. The expression `static' refers to time-independent properties. {We treat the ground state energy per site in Sect.\ \ref{subsec:gse}, the magnetization in Sect.\ \ref{subsec:magnetization} and the  momentum-integrated spectral weight in Sect.\ \ref{subsec:spectral_weight} and 
the momentum-resolved static structure factor  in Sect.\ \ref{subsec:eq_time_sf}.}

\subsection{Ground state energy}
\label{subsec:gse}

Due to the only linearly growing number of string operators, we are able to obtain the ground state energy per site up to {order 256}. Higher orders do not improve the results significantly
so that we restrict ourselves to orders up to 256.

\begin{figure}
\centering
\includegraphics[width=\columnwidth]{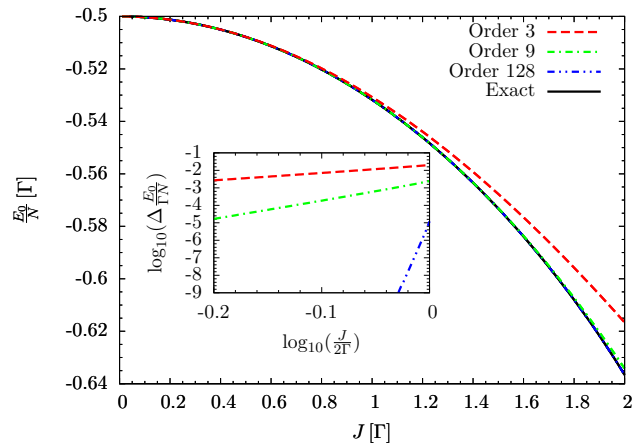}
\caption{(Color online) Ground state energy per site as function of $J$. 
Comparison of the exact result to CUT results in various  orders of $x=J/(2\Gamma)$.
The inset shows the absolut difference between the exact result and the 
CUT calculation on a logarithmic scale. The critical point is located at $\log_{10}\left(\frac{J}{2\Gamma}\right) = 0$.}
\label{image:GSE_Exact_vs_cut_differntorders}
\end{figure}

Figure \ref{image:GSE_Exact_vs_cut_differntorders} compares the exact result for the ground state energy
per site to {CUT results in}
various orders in $x$. As expected, the accuracy increases for increasing
order. Even close to the critical point the CUT result of order 128 and the exact results can barely
be separated. The inset shows the {deviations from the exact results}. 
{On} the logarithmic scale, straight
lines indicate power laws for these deviations {as expected} in a perturbatively
controlled approach. We checked that the slopes of the lines correspond to the order of 
calculation by fitting the deviations to power laws, see Tab.\ \ref{tab.PowerLaw_GSE}.

\begin{table}[htb]
\centering
\begin{tabular}{|c|c|c|}
\hline
\text{Order} & \text{Exponent} & \text{Fitting Error} \\
\hline
9 	&	10  &  $\pm$ 3 \\
16  &   21  &  $\pm$ 4 \\
32  &   33  &  $\pm$ 3 \\
64	& 	72  &  $\pm$ 6 \\
128	&	132 &  $\pm$ 5 \\
\hline             
\end{tabular}
\caption{Exponents of the power laws for the deviation of the ground state energy obtained by numerical
fits.}
\label{tab.PowerLaw_GSE}
\end{table}

From the inset we also read off that the deepCUT in order 128 calculates the ground state energy per site {correctly to the fifth digit}, even at the critical point. The calculation in order 256 is 
not shown in the graphs because it would be indistinguishable from the other curves. 
It improves the result in order 128  at the critical point by about one {digit}.

\subsection{Transverse magnetization}
\label{subsec:magnetization}

Next, we examine the transverse magnetization
\begin{align}
M_z = - \frac{1}{N} \sum\limits_{j} \bra{g} \sigma^z_j \ket{g}.
\end{align}
Here $\ket{g}$ denotes the ground state of the Hamiltonian which is {mapped to}
the zero-particle state by the CUT.
Note that the expression `transverse' refers to the direction of the external field, which is 
{the z-axis in our model}. In the limit $J \rightarrow 0$, all spins are aligned along the external field, so that $M_z = 1$ holds. 

\begin{figure}
\centering
\includegraphics[width=0.5\textwidth]{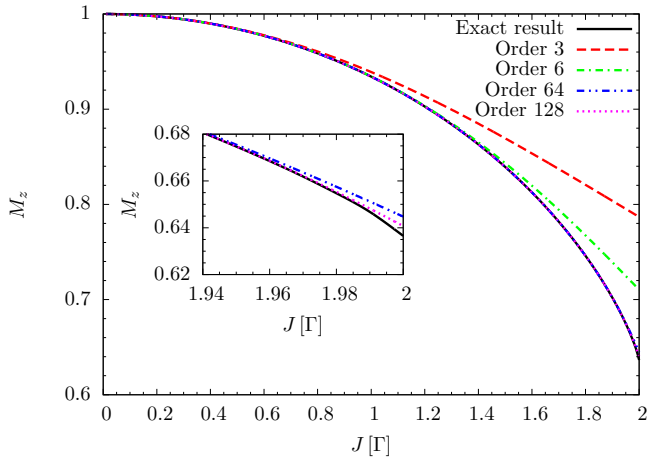}
\caption{(Color Online) Transverse magnetization as a function of $J$. Comparison of the exact result with the CUT calculation. The inset shows a close view on the critical point $J=2{\Gamma}$.}
\label{image:MZ_exact_vs_cut_differentorders}
\end{figure}

For $J>0$, the spins are perturbed by the antiferromagnetic interaction which reduces the transverse magnetization. To obtain $M_z$, we transform the observable $\sigma^z_j$ by the continuous unitary transformation to an effective observable. The operator of the transverse magnetization can be expressed by a string operator
\begin{align}
\sum\limits_{j} \sigma^z_j  (l=0) = T_0.
\end{align}
Due to this identity and the fact that the string algebra is closed under commutation,
we know that the final effective observable can be written as a linear combination of string operators
\begin{subequations}
\begin{align}
\sigma^z (l)  
	&= o_0(l) T_0 + \sum\limits_{n} \left[ o_n^{+-}(l) \left(T_n^{+-} + \mathrm{h.c.} \right) \right. \\
	&+ \left.	o_n^{++}(l) \left(T_n^{++} + \mathrm{h.c.} \right) \right] \nonumber .
\end{align}
\end{subequations}
None of the coefficients $o_n^{+-}$ and $o_n^{++}$ contribute to the vacuum expectation value. But
they can not be omitted during the flow of the observable. 
The transverse magnetization after the CUT reads
\begin{align}
M_z  = o_0(\infty) .
\end{align}
Due to this simple form of the observable very high orders can be reached again.
The transverse magnetization calculated by the CUT  is compared to the exact result in Fig.\ \ref{image:MZ_exact_vs_cut_differentorders}. As expected the results improve upon increasing order. The largest error occurs at the critical point where the transverse magnetization displays a singularity.

\subsection{Spectral weight}
\label{subsec:spectral_weight}

In this section, we discuss the CUT results for the {momentum-integrated}
quasi-particle weight in the two diagonal channels $S^{xx}$ and $S^{yy}$.
We use the CUT framework to calculate effective observables which renders  an easy evaluation for the spectral weight {possible} in various quasi-particle channels, see also Ref.\ \onlinecite{PhysRevLett.92.027204}. The total spectral weight can be split according to
\begin{align}
I^{\alpha \alpha} = 
\frac{1}{N} \sum\limits_l \left\langle \sigma_l^\alpha \sigma_l^\alpha \right\rangle = 
I_1 + I_2 + I_3 + \dots ,
\end{align}
where $I_n$ denotes the {weight in the} channel with $n$ quasi-particles in the system. 
Introducing the CUT framework results in
\begin{align}
I^{\alpha \alpha}_n =  \bra{0} \sigma_{n, \mathrm{eff}}^{\alpha} \sigma_{n, \mathrm{eff}}^{\alpha, \dagger} \ket{0},
\end{align}
where $\sigma_{n, \mathrm{eff}}^{\alpha}$ denotes the part of the effective observable {which} annihilates $n$ quasi-particles. Since $\sigma^x$ and $\sigma^y$ create an odd number of spin flips, i.e., quasi-particles, and since the generator of the CUT preserves the parity of an observable, 
$S^{xx}$ and $S^{yy}$ consist of $1,3,5, \dots$ quasi-particle contributions. 
With the help of the sum rule $I^{{\alpha\alpha}}=1$, 
we can also check if our results are still valid for large {values of} $x$. 

We emphasize that the local observables $\sigma_l^x$ and $\sigma_l^y$ transform into non-local operators under the Jordan-Wigner transformation which act on an infinite number of sites. Therefore, no easy analysis of these observables {is possible} in fermionic terms. {An explicit evaluation} requires either analytical mappings which enable an evaluation in the scaling region 
\cite{Vaidya19781} or extensive numerics {in terms of }
Pfaffians {whose} dimensions grow linearly with the spatial range of the correlation \cite{StolzeKrones}.

The problem of an infinite number of  operators is avoided in the string operator basis \eqref{eq:string_basis}.
But the calculation remains cumbersome because the observables are \emph{not}
 part of the string algebra
and thus the structure of the effective observables is more complicated. This complicated structure prevents us from achieving very high orders, because the number of representatives to be tracked grows exponentially on increasing order. We are able to obtain results up to order $38$. Then the computational effort reaches its limit in the present {implementation} because the contributions to the differential equation system take more than $8$ GB of memory and the number of operators to track is larger than $7$ million.

\begin{figure}
\centering
\includegraphics[width=0.5\textwidth]{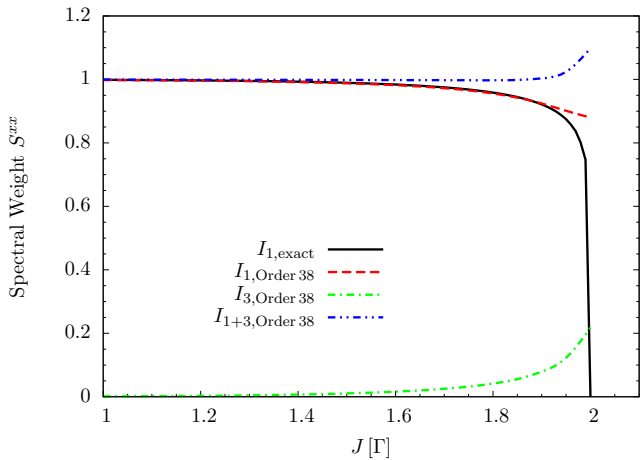}
\caption{(Color Online) One-particle and three-particle spectral weight as function of the parameter $J$. Comparison of the exact expression for the one-particle weight \eqref{S1xx} with the CUT results.}
\label{image:DSF_xx_1p_exact_vs_cut_sum}
\end{figure}

First, we address $S^{xx}$ as function of $J$ {for which results are}
depicted in Fig.\ \ref{image:DSF_xx_1p_exact_vs_cut_sum}. The CUT results are compared to the exact results from Ref.\ \onlinecite{HamerTFIM}. 
The one-particle contribution shows a very sharp drop for $J \to 2\Gamma$ with a singularity at the critical point. The CUT agrees very well with the exact results as long as the order of calculation is below the correlation length. Recall that the order is proportional to the range of the physical
processes included in the calculation.
For the calculation of effective observables within the string algebra
this was no problem because large orders $>100$ could be achieved. 
But for the longitudinal correlations, we obtain only order 38 so that particularly
sharp edges such as the one in the one-particle spectral weight are not captured. 
But the agreement improves on increasing order.

The spectral weight of the three-particle channel increases on approaching the critical point. Hence, spectral weight is transferred from the one-particle channel to higher quasi-particle channels. 
The one- and three-particle terms are the dominant contributions to the total spectral weight for the parameters investigated.
But for $J>1.9$, the sum rule starts being violated in the CUT calculation.
{We attribute this violation} 
to the calculation in finite order. It appears that the CUT calculation overestimates the 
one-particle contributions close to the critical point.

\begin{figure}
\centering
\includegraphics[width=0.5\textwidth]{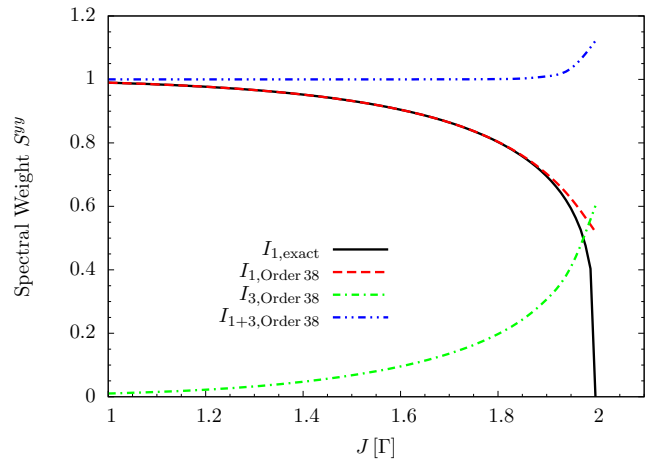}
\caption{(Color online) One-particle and three-particle spectral weight as function of the parameter $J$. Comparison of the exact expression for the one-particle weight \eqref{S1yy} with the CUT results.}
\label{image:DSF_yy_1p_exact_vs_cut_sum}
\end{figure}

Next, we investigate $S^{yy}$.
This correlation is depicted in Fig.\ \ref{image:DSF_yy_1p_exact_vs_cut_sum} in comparison to the exact result. Again, the one-particle contributions also vanish for $J \rightarrow 2\Gamma$. But the edge at the critical point is by far not as sharp as in $S^{xx}$ because more spectral weight is transferred to higher quasi-particle spaces for lower parameters $J$. The sum rule is again violated for $J>1.9\Gamma$ due to finite order errors.

\subsection{Equal-time structure factor}
\label{subsec:eq_time_sf}

\begin{figure}
\centering
\includegraphics[width=0.5\textwidth]{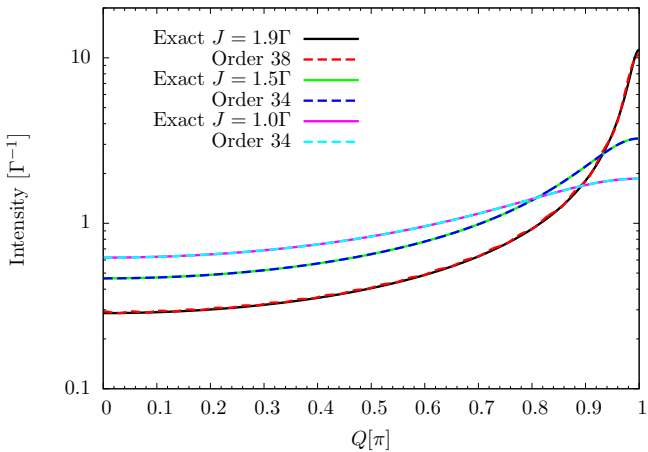}
\caption{(Color Online) One-particle equal time structure factor $S_1^{xx}(Q)$ for the parameters $J=\Gamma$, $J=1.5\Gamma$ and $J=1.9\Gamma$. Comparison of the exact expression \eqref{S1xx} for the one-particle weight  with the CUT results. }
\label{image:DSF_xx_1p_exact_vs_cut_multij}
\end{figure}

{The momentum-resolved equal-time structure factor contains more information
so that it is another interesting quantity}
\begin{align}
S^{\alpha \alpha}(Q) = \frac{1}{N} \sum\limits_{l, l'} e^{-i Q(l-l')} \langle \sigma_l^\alpha \sigma_{l'}^\alpha \rangle .
\end{align}
For a single quasi-particle it is directly connected to the full DSF by Eq.\ \eqref{eq:equal_time_dsf}
because there is no mixing between different quasi-particle spaces \cite{knett01b}. 
Our first focus is $S_1^{xx}(Q)$. Within the CUT framework, this quantity can be computed from the effective observable $\sigma^x_{j,\mathrm{eff}}$ by Fourier transformation of the terms that exactly create one particle
\begin{align}
S^{\alpha \alpha}_n(Q) =  \bra{0} \sigma_{n, \mathrm{eff}}^{\alpha}(-Q) \sigma_{n, \mathrm{eff}}^{\alpha, \dagger}(Q) \ket{0} 
\end{align}
where $n$ stands for the number of quasi-particles involved and $\alpha$ may take
the values $x$ or $y$.
In Fig.\ \ref{image:DSF_xx_1p_exact_vs_cut_multij} we compare the CUT results to the exact expression 
\eqref{S1xx} for the parameters $J=\Gamma$, $J=1.5\Gamma$ and $J=1.9\Gamma$. 
The agreement is very impressive though it worsens upon approaching the critical point.
For $J=\Gamma$ the DSF is essentially converged and the absolute errors are below $10^{-10} \Gamma^{-1}$.  For $J=1.9\Gamma$ the error is below $10^{-3} \Gamma^{-1}$ for $Q < \pi/2$. 
For $Q > \pi/2$ the absolute error rises up to $10^{-1} \Gamma^{-1}$.

A closer analysis reveals that the largest absolute error occurs 
{for all parameters at the wave vector $Q=\pi$}.
The DSF diverges at this point for $J \rightarrow 2\Gamma$. The relative error (not shown in the graphs) 
remains fairly constant over the whole Brillouin zone. Consequently our results differ from the
exact ones found by Hamer et al.\ \cite{HamerTFIM} by only about $1\%$ even close to the critical point at $J=1.9\Gamma$. 

\begin{figure}
\centering
\includegraphics[width=0.5\textwidth]{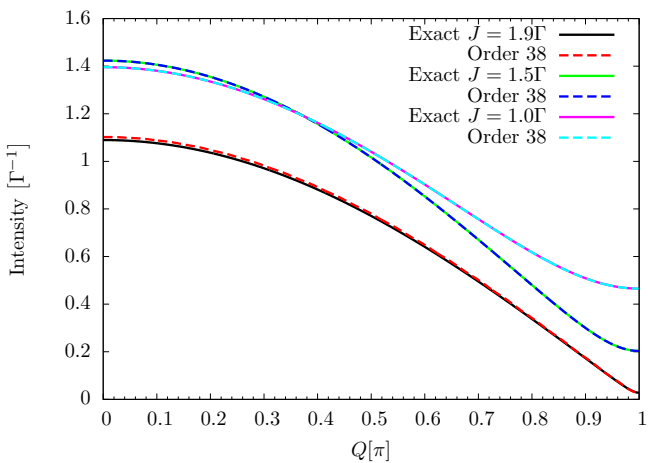}
\caption{(Color online) One-particle equal time structure factor $S_1^{yy}(Q)$ for the parameters $J=\Gamma$, $J=1.5\Gamma$ and $J=1.9\Gamma$. Comparison of the exact expression \eqref{S1yy} with the CUT results. }
\label{image:DSF_yy_1p_exact_vs_cut_multij}
\end{figure}

Finally, we consider $S_1^{yy}(Q)$.
In Fig.\ \ref{image:DSF_yy_1p_exact_vs_cut_multij} we compare the CUT results to the exact expression for the parameters $J=\Gamma$, $J=1.5\Gamma$ and $J=1.9\Gamma$. 
For $J=\Gamma$ the DSF is essentially converged and the absolute errors are below $10^{-10} \Gamma^{-1}$. This changes for rising parameter $J$. For $J=1.9\Gamma$ the error is below $10^{-2} \Gamma^{-1}$ for $Q < \pi/2$. For $Q > \pi/2$ the absolute error {remains}  below $10^{-3} \Gamma^{-1}$.
In contrast to the $S^{xx}$ channel, the lowest absolute error is  located in the  $S^{yy}$ channel
at $Q \approx \pi$. This is can be easily understood because $S_1^{yy}(Q=\pi)$ constitutes a local minimum for 
all parameters $J$. Again, the relative error (not shown) 
remains essentially constant over the whole Brillouin zone.

\section{Dynamic properties} 
\label{sec:dynamic_results}

In this section, we evaluate and present the dynamic results for the transverse field Ising model.
Here, `dynamic' refers to frequency dependent quantities. {We deal with 
the dispersion in Sect.\ \ref{subsec:dispersion} and with the DSF in general in Sect.\ \ref{subsec:dsf}
and its different channels in Sects.\  \ref{subsec:dsf_zz} ($S^{zz}$), \ref{subsec:dsf_xx} ($S^{xx}$) and \ref{subsec:dsf_yy} ($S^{yy}$).} 

The {general} DSF is an important quantity because it is directly measurable in scattering experiments. Furthermore, dynamic correlations strongly depend on the model under study and often exhibit features which reveal the microscopic interactions in the Hamiltonian. Despite the fact that the TFIM is integrable, the calculation of dynamic correlations remains a difficult and complex problem.

\subsection{Dispersion}
\label{subsec:dispersion}

\begin{figure}
\centering
\includegraphics[width=0.5\textwidth]{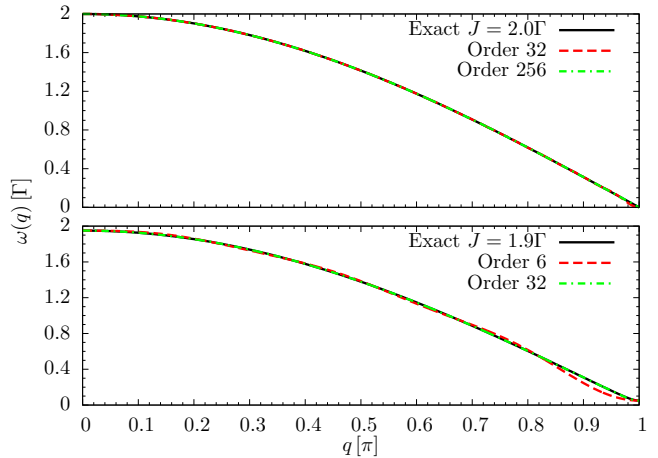}
\caption{(Color online) Energy dispersion for $J=2.0\Gamma$ (top) and $J=1.9\Gamma$ (bottom). Comparison of the the CUT calculations to the exact results.}
\label{image:Dispersion_exact_vs_cut_differentorders_j1920}
\end{figure}

As before we were able to reach order 256 for the CUT calculation of the Hamiltonian.
In particular, we obtain the hopping matrix elements up to a range of 256. 
For small parameters $J$, a low order calculation is sufficient to achieve a good agreement with the exact result. Closer to the critical point this changes distinctly, see Fig.\ \ref{image:Dispersion_exact_vs_cut_differentorders_j1920}, which makes higher order calculations necessary. 
For $J=1.9\Gamma$ the absolute error of the result in order 6 is below $0.01 \Gamma$ for $q < \pi/2$ 
and below $0.1 \Gamma$ for $q > \pi/2$. For the order 32 result, it is below $10^{-4} \Gamma$ for $q < \pi/2$ and below $10^{-3} \Gamma$ for $q > \pi/2$.
For $J=2\Gamma$, the error of the order 32 result is below $10^{-3} \Gamma$ for $q < \pi/2$ and below $10^{-1} \Gamma$ for $q > \pi/2$. For the order 256 result,
 it is below $10^{-5} \Gamma$ for $q < \pi/2$ and below $10^{-3} \Gamma$ for $q > \pi/2$.

This behavior is expected because the excitations become more and more dispersive on increasing $J$.  {Consequently, hopping processes over larger and larger distances become more important.} To include these physical processes we need higher orders because the maximum range we can describe directly corresponds to the order of calculation (for lattice constant equal to unity). {Directly at the critical point the energy gap closes and the correlation length diverges concomitantly.}

The calculation of the dispersion is worst in the vicinity of the critical wave vector $q=\pi$. We stress, however, that the value directly at $q=\pi$, the energy gap of the TFIM,
 is calculated exactly up to numerical errors below $10^{-10}\Gamma$. 
This is an accidental result because the energy gap happens to be
 a linear function of $J$ so that it is captured correctly by any perturbative approach in linear order and beyond, compare Eq.\ \eqref{eq:gap_exact}.

\subsection{Dynamic structure factor}
\label{subsec:dsf}

The DSF at $T=0$ is linked to the imaginary part of the retarded Green function by the fluctuation-dissipation theorem at zero temperature \cite{Chandler}
\begin{align} 
\label{Eq.ConnectionSandG}
S^{\alpha \alpha}(\omega, Q) = - \frac{1}{\pi} \mathrm{Im} G^{\alpha \alpha}(\omega, Q).
\end{align}
At $T=0$, it is useful to write this Green function for $\omega>0$ as a resolvent
\begin{align} 
\label{Eq.Resolvente}
G^{\alpha \alpha}(\omega, Q) &=  \langle g | \sigma^\alpha(-Q) \nonumber \\
& \frac{1}{\omega-(H(Q)-E_0)+i0^+} \sigma^\alpha(Q) | g \rangle 
\end{align}
where $E_0$ is the ground state energy and 
\begin{align}
\sigma^\alpha(Q) = \frac{1}{\sqrt{N}} \sum\limits_{l} e^{i Q l} \sigma^\alpha_l
\end{align}
is the Fourier transformed spin operator $\sigma_l^\alpha$.

In the CUT framework, we replace all operators by the effective operators and the ground state by the zero-particle state, i.e., the vacuum of quasi-particles
\begin{align} 
\label{eq.DSF_CUT}
G^{\alpha \alpha}(\omega, Q) &= \langle 0 | S_\mathrm{eff}^\alpha(-Q) \nonumber \\
						&\frac{1}{\omega-(H_\mathrm{eff}(Q)-E_0)+i0^+}  S_\mathrm{eff}^\alpha(Q)  | 0 \rangle . 
\end{align}
We evaluate the resolvent in Eq.\ \eqref{eq.DSF_CUT} by means of a Lanczos tridiagonalization yielding
 a continued fraction representation of the resolvent \cite{petti85, mueller} 
\begin{align} 
\label{eq.conti_frac}
G^{\alpha \alpha}(\omega, Q) = \frac{b_0^2}{\omega - a_0 - \frac{b_1^2}{\omega - a_1 - \frac{b_2^2}{\ddots} } } ,
\end{align}
where the coefficients $a_n$ and $b_n$ are the matrix elements of the tridiagonal matrix of the effective Hamiltonian. We refer the reader to Appendices \ref{app:c} and \ref{app:d} where we explicitly {calculate $S_\mathrm{eff}^\alpha(Q)  | 0 \rangle$} as well as the action of the effective Hamiltonian {for the Lanczos tridiagonalization}. The continued fraction is terminated by  a
standard square-root terminator. This is appropriate for square root singularities at the band-edges.

\begin{figure}
\centering
\includegraphics[width=0.45\textwidth]{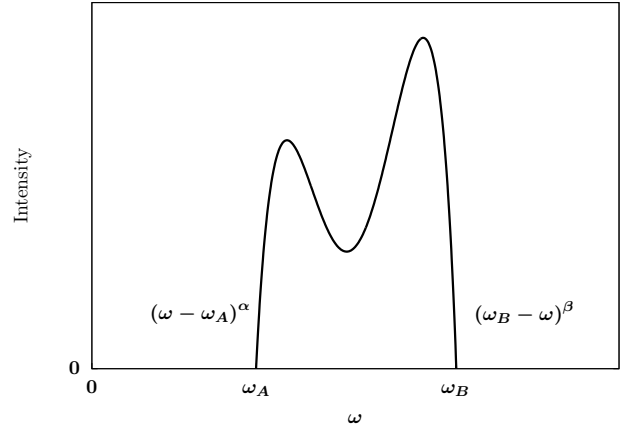}
\caption{Qualitative sketch of the band-edge singularities in the DSF.}
\label{image:BandEdge_Example}
\end{figure}

Another piece of information that can be obtained from the sequences $\left\lbrace a_n \right\rbrace$ and $\left\lbrace b_n \right\rbrace$ are the exponents $\alpha$ and $\beta$
of the band-edge singularities, see Fig.\ \ref{image:BandEdge_Example}. 
They are directly connected to the asymptotics \cite{petti85}
\begin{subequations}
\begin{align}
a_n &= a_\infty + b_\infty \frac{\beta^2-\alpha^2}{2n^2} + \mathcal{O}\left(\frac{1}{n^3}\right) \label{Eq.a_conv} \\
b_n &= b_\infty + b_\infty \frac{1 - 2\alpha^2 - 2\beta^2}{8n^2} + \mathcal{O}\left(\frac{1}{n^3}\right) .  \label{Eq.b_conv}
\end{align}
\end{subequations}
These relations allow us to obtain the band-edge singularities up to their signs by fitting 
\begin{align}
f(n) = C + \frac{D}{n^2}
\end{align}
to the continued fraction coefficients.

For two massive hardcore particles without interaction and with finite range hopping in one dimension the band-edge singularities are known to be $\alpha = \beta = 1/2$, see for instance 
Refs.\ \onlinecite{PhysRevB.54.R9624, PhysRevB.58.2900}. We expect this behavior also to be true in the case of the TFIM because there is no interaction  in the exact solution. 
In this case the relations \eqref{Eq.a_conv} and \eqref{Eq.b_conv} yield
\begin{subequations}
\begin{align}
a_n &= a_\infty + \mathcal{O}\left(\frac{1}{n^3}\right) \\
b_n &= b_\infty + \mathcal{O}\left(\frac{1}{n^3}\right) .
\end{align}
\end{subequations}
We confirm this behavior in the two-particle case {of the $S^{zz}$ channel in Sect.\ \ref{subsec:dsf_zz}.}

\subsection{$S^{zz}$ channel}
\label{subsec:dsf_zz}

Because the observable $\sigma^z$ stays local in the Jordan-Wigner representation of the TFIM the DSF
in the $S^{zz}$ channel can be obtained analytically, see Eq.\ \eqref{Eq.DSF_zz_exact}. 
The  DSF {in the $zz$ channel} results from the two-particle continuum. Even for large parameters $J$, no weight is shifted towards four or more particle spaces. 
All dynamics induced by this observable is captured in the two-particle sector. 

We emphasize that this fact holds as well in in the CUT treatment formulated in
terms of the string operator algebra. The corresponding local operator $\sigma_j^z = O_{j,0}$ is {element} of the string algebra so that its effective observable after the CUT consists of a linear combination of string operators
\begin{align}
\sigma_{j,\mathrm{eff}}^z &= \sum\limits_{d} o_{j+d} O_{j+d,0} 
\\ 
+& \sum\limits_{d,n} \left[ o_{j+d,n}^{+-} ( O_{j+d,n}^{+-} 
+ \mathrm{h.c.} ) + o_{j+d,n}^{++} ( O_{j+d,n}^{++} + \mathrm{h.c.} \right]) \nonumber 
\end{align}
where the maximum range $n$ is limited by the order of the calculation. {We stress again that the operators $O_{j+d,0}$ and $O_{j+d,n}^{+-}$ do not create any excitations, while the operators $O_{j+d,n}^{++}$ create exactly two excitations. Thus, the vector 
$S_\mathrm{eff}^\alpha(Q)  | 0 \rangle$ is only element of the zero- and of the two-particle Hilbert space.} The same holds in the fermionic picture, where the operator $\sigma_i^z$ is a local density term which at most creates two fermionic excitations after the Bogoliubov diagonalization.

Concomitantly,
very high orders can be reached also in the transformation of the local observable. 
Because we transform a non translational-invariant operator, we have to consider the positions 
$j+d$ and the starting site $j$ so that the number of terms increases quadratically with the order.
But we are still able to achieve an order of $128$.

\begin{figure}
\centering
\includegraphics[width=0.5\textwidth]{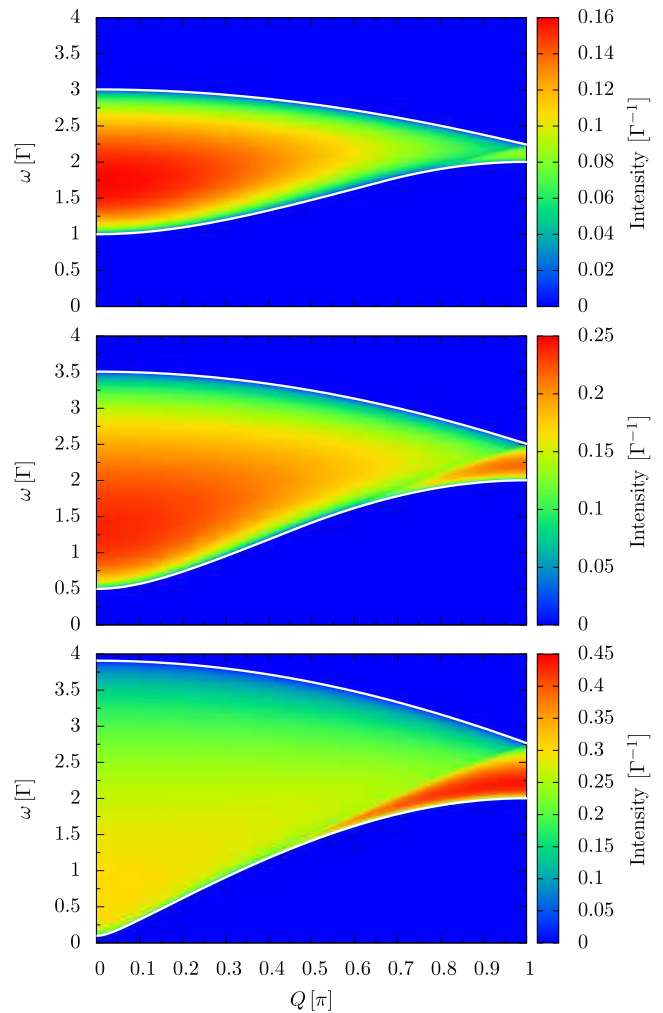}
\caption{(Color online) The DSF $S^{zz}{(\omega,Q)}$ for the parameters $J=\Gamma$ (top), $J=1.5\Gamma$ (center) and $J=1.9\Gamma$ (bottom). The maximum range for the Lanczos algorithm is $d_\mathrm{max}=1000$ sites, the continued fraction was evaluated to a depth of $50$ and then terminated by the square root terminator. The color indicates the spectral density, see legend to the right. The upper and lower edge of the two-particle continuum are indicated by white lines.}
\label{image:DSF_zz_exact_j101519}
\end{figure}

False color plots of the DSF obtained in this way are shown in Fig.\ \ref{image:DSF_zz_exact_j101519} in 
order 128. The two-particle continuum is depicted in dependence of the total momentum $Q$ and the energy $\omega$. The overall intensity rises for larger parameters $J$. This stems from the decrease of the transverse
magnetization which induces a shift of spectral weight from the zero-particle channel
to the two-particle channel. 
Furthermore, we see that for small parameters $J/(2\Gamma)$
most of the weight is concentrated in the region $Q<\pi/2$ while the opposite behavior occurs 
for larger parameters $x$. 
Note the singularity inside the continuum on the right side of the Brillouin zone that separates two regions with low and high spectral weight, see the case $J=1.9\Gamma$. 
Knowing the exact expression \eqref{Eq.DSF_zz_exact} we can explain this singularity as van Hove singularity in the two-particle density-of-states. The two-particle energy $\omega(Q/2+q) + \omega(Q/2-q)$ displays a \emph{local} maximum  besides the global extrema as function of $q$, if $Q$  and $J$ are large enough. 

\begin{figure}
\centering
\includegraphics[width=0.5\textwidth]{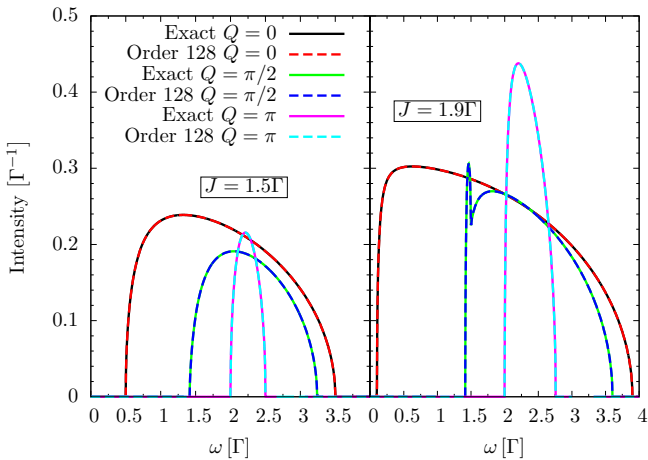}
\caption{(Color online) DSF $S^{zz}{(\omega,Q)}$ for the parameters $J=1.5\Gamma$ (left) and $J=1.9\Gamma$ (right) for three total momenta $Q$. The maximum range for the Lanczos algorithm is $d_\mathrm{max}=4000$ sites, the continued fraction was evaluated to a depth of $100$ and then terminated by the square root terminator.}
\label{image:DSF_zz_exact_vs_cut_j1519_diffq}
\end{figure}

We want to investigate more profoundly how the CUT calculation differs from the exact calculation by examining the DSF for fixed parameters $J$ and total momentum $Q$. Fig.\ \ref{image:DSF_zz_exact_vs_cut_j1519_diffq}
shows $S^{zz}{(\omega,Q)}$ for $J=1.5\Gamma$ and $J=1.9\Gamma$ and for the momenta $Q=0$, $Q=\pi/2$ and $Q=\pi$ calculated by the CUT in order 128 in comparison to the exact result. Note the {excellent} agreement for all parameters and momenta. The form of the DSF is very close to a half ellipse for low values of $J$ because the continued fraction coefficients converge very quickly towards their final values $a_\infty$ and $b_\infty$.
For large $J$ more spectral weight is concentrated at the lower band-edge which we attribute to a complex interplay between momentum and energy conservation. For the parameters $J=1.9$ and $Q=\pi/2$, one clearly sees
the singularity inside the continuum of the DSF which is the above mentioned
van Hove singularity from a local maximum. 

A detailed analysis shows that the error is lower in the middle of the continuum than at the band-edge singularities. This is expected due to the strong change of the DSF at the edges. 
On average, the error is below $10^{-6} \Gamma^{-1}$ even for large parameters $J$. At the band-edges the error can rise up to $10^{-3} \Gamma^{-1}$. We presume that the errors are mainly due to inaccuracies in the Lanczos tridiagonalization and due to the limited maximum range in the transformation of the observable by the CUT. Nonetheless the errors are still very small and {justify} our approach.

\begin{figure}
\centering
\includegraphics[width=0.5\textwidth]{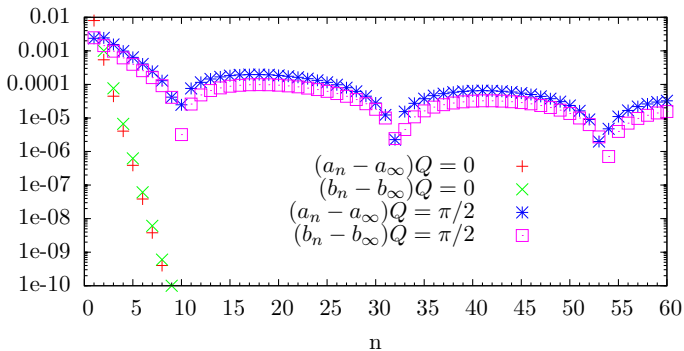}
\caption{(Color online) Absolute difference between the continued fraction coefficients and their final values for the case $J=1.5\Gamma$ and total momenta $Q=0$ and $Q=\pi/2$.}
\label{image:DSF_zz_fraction_j15_diffq}
\end{figure}

Next, we investigate how the continued fraction coefficients approach their limiting values. 
Thereby, we estimate
the exponents of the band-edge singularities  according to Eqs.\ \eqref{Eq.a_conv} and  \eqref{Eq.b_conv}.
The continued fraction coefficients for the case $J=1.5\Gamma$ and total momenta $Q=0$ and $Q=\pi/2$ are shown in Fig.\ \ref{image:DSF_zz_fraction_j15_diffq}. The coefficients for the case $Q=0$ approach their limit exponentially. Therefore we know by Eqs.\ \eqref{Eq.a_conv} and \eqref{Eq.b_conv} that both exponents take the value $1/2$. 

For the case $Q=\pi/2$, the coefficients do not converge so rapidly. We fit them for this case versus 
$1/n^2$ to check if they display terms  in $\mathcal{O}(1/n^2)$. Both coefficients oscillate around the final value which can not be described by Eqs.\ \eqref{Eq.a_conv} and \eqref{Eq.b_conv}. This again indicates that the exponents at the band-edges are $1/2$. We also checked other momenta and they support the assumption that all exponents are  $1/2$ for the two-particle case as it has to be according to the fermionic analytical results.
Thus, these findings corroborate the validity of our approach and analysis.

\subsection{$S^{xx}$ channel}
\label{subsec:dsf_xx}

In Ref.\ \onlinecite{HamerTFIM} Hamer et al.\ derived an analytic expression for the one-particle contribution
for the longitudinal structure factors. 
To our knowledge no data is available in the literature for higher quasi-particle contributions 
away from the scaling region \cite{Vaidya19781}. Here our approach is able to provide complementary quantitative knowledge.

\begin{figure}
\centering
\includegraphics[width=0.5\textwidth]{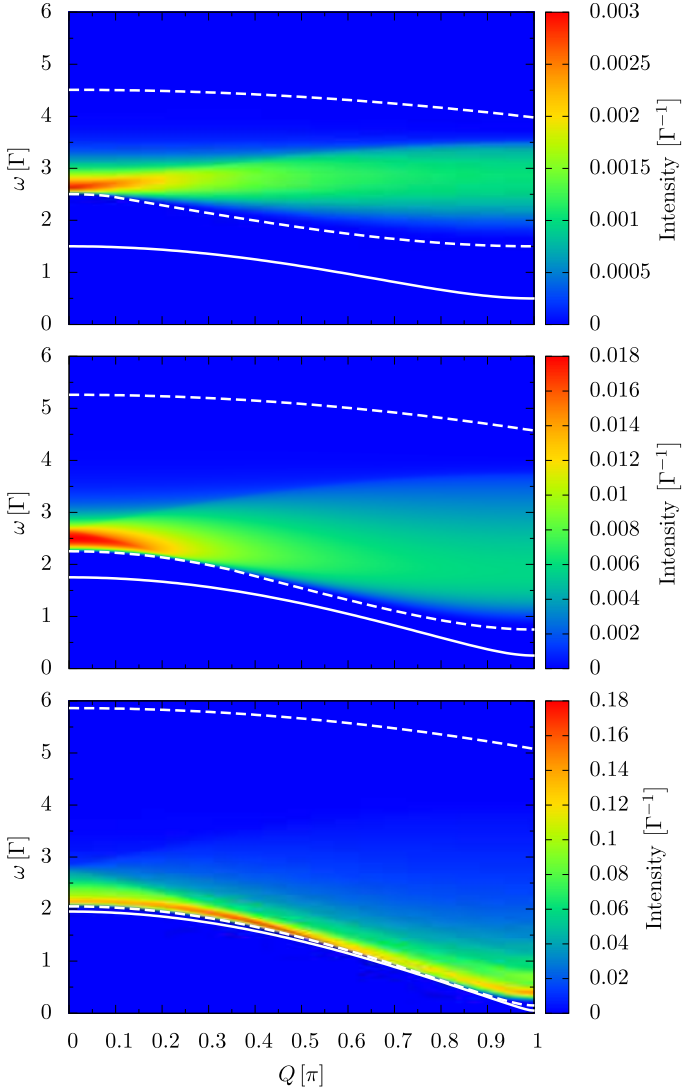}
\caption{(Color online) The DSF $S_3^{xx}{(\omega,Q)}$
 for the parameters $J=\Gamma$ (top), $J=1.5\Gamma$ (center) and $J=1.9\Gamma$ (bottom). The maximum range for the Lanczos algorithm is $d_\mathrm{max}=100$ sites, the continued fraction was evaluated to a depth of $50$ and then terminated by the square root terminator. The color indicates the spectral density, see legend to the right. The dispersion is indicated by the white solid line. The upper and lower edge of the three-particle continuum are indicated by white dashed lines. All results are computed in order 38.}
\label{image:DSF_xx_3p_j101519}
\end{figure}

Similar to the two-particle case $S^{zz}{(\omega,Q)}$, the three-particle case $S_3^{xx}{(\omega,Q)}$ consists of a continuum of states. 
We are limited to a maximum order $38$ due to the complicated structure of the local observable $\sigma^x$ which is not part of the closed string algebra.
Overview plots for the DSF obtained by the CUT are found in Fig.\ \ref{image:DSF_xx_3p_j101519}. In these plots  the three-particle continuum is depicted in dependence of total momentum $Q$ and the energy $\omega$. 

The total weight rises on increasing  $J$ because spectral weight is transferred from the one-particle sector to the higher quasi-particle channels.  The same qualitative behavior is observed for dimerized spin chains and spin ladders, and related systems \cite{GoetzSpinChain, GoetzSpinLadder,IPA_CuCl}.
In addition, we notice that most of the spectral weight is concentrated at momenta $Q<\pi/2$ for small parameters $J$. The weight slowly shifts for growing $J$ similar to the $S^{zz}$ case.
For $J=1.9 \Gamma$, most of the spectral weight is concentrated rather strongly at the lower band-edge of the continuum. The same tendency was found in the $S^{zz}$ case as well. 
Still the shape of the DSF  in the  $S^{xx}$ case differs strongly from a semi-ellipse in 
contrast to the $S^{zz}$ DSF. 

\begin{figure}
\centering
\includegraphics[width=0.5\textwidth]{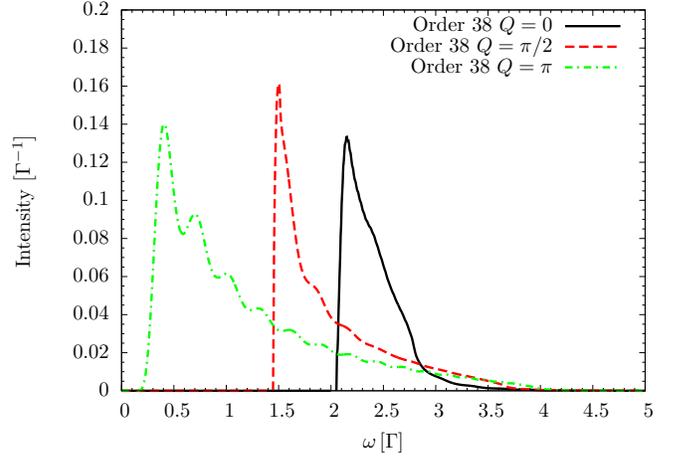}
\caption{(Color online) DSF $S_3^{xx}{(\omega,Q)}$ for the parameter $J=1.9\Gamma$ for three chosen total momenta $Q$. The maximum range for the Lanczos algorithm is $d_\mathrm{max}=300$ sites, the continued fraction was evaluated to a depth of $100$ and then terminated by the square root terminator.}
\label{image:DSF_xx_3p_j19_q_fixed}
\end{figure}

A more quantitative investigation is shown in Fig.\ \ref{image:DSF_xx_3p_j19_q_fixed} where $S_3^{xx}$ is plotted for  $J=1.9\Gamma$ and momenta $Q=0$, $Q=\pi/2$ and $Q=\pi$. It is confirmed that most of the spectral weight is concentrated at the lower band-edge for large values of $J$. For $Q=\pi$, a strong wiggling occurs
which is to be attributed to the errors due to the  calculation in finite order.

\begin{figure}
\centering
\includegraphics[width=0.5\textwidth]{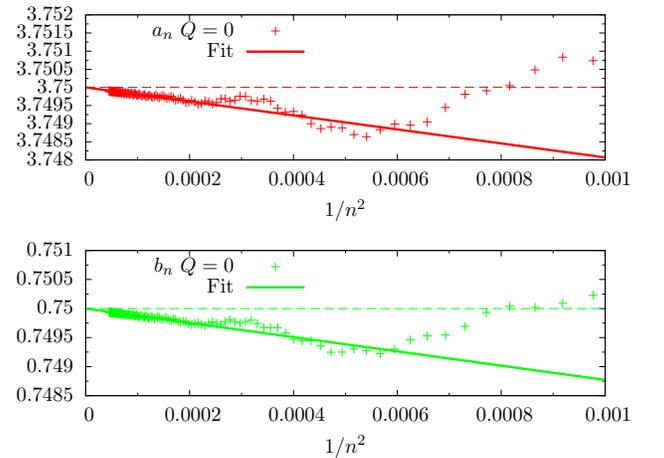}
\caption{(Color online) Continued fraction coefficients for the case $J=1.5\Gamma$ and total momentum $Q=0$. The upper panel shows the coefficients $a_n$ and the lower panel shows the coefficients $b_n$. The limit values are indicated by horizontal lines. The red/green lines indicated linear fits in $1/n^2$; note the scale of the $x$-axis.}
\label{image:DSF_xx_fraction_overn_q0}
\end{figure}

Next, we investigate the band-edge singularities in the three-particle case $S_3^{xx}{(\omega,Q)}$. As in the two-particle case, we use the relations \eqref{Eq.a_conv} and \eqref{Eq.a_conv} by fitting 
a $1/n^2$ power law to the continued fraction coefficients.
An example is shown in Fig.\ \ref{image:DSF_xx_fraction_overn_q0}

 In contrast to the two-particle case, no exponential approach towards the limit values occurs. 
 For all momenta, both $a_n$ and $b_n$ show a behavior proportional to $1/n^2$. We stress that the $\mathcal{O}(1/n^3)$ terms are significant up to $1/n^2 \approx 0.0002 \, \Rightarrow n \approx 70$.
The values for the band-edge singularities obtained from the fits are shown in Tab.\ \ref{tab.BandEdge_Sing_xx}.

\begin{table}
\centering
 \begin{tabular}{|c|c|c|}
\hline
$Q$ & $\alpha$ & $\beta$ \\
\hline
0	   &	2.5 $\pm$ 0.3 & 1.0 $\pm$ 0.2     \\
$\pi/2$ &   3.0 $\pm$ 0.2   & 2.7 $\pm$ 0.2	  \\
$\pi$  &    3.0 $\pm$ 0.2   & 2.8  $\pm$ 0.1      \\   
\hline  
\end{tabular}
\caption{Exponents for the band-edge singularities of $S_3^{xx}{(\omega,Q)}$ for $J=1.5\Gamma$. The errors are determined from the fits using the Levenberg-Marquardt algorithm \cite{Levenberg, Marquardt}.}
\label{tab.BandEdge_Sing_xx}
\end{table} 

For $Q=\pi$ and $Q=\pi/2$, both exponents are close to  $3$ while for $Q=0$ the exponents differ and 
we {deduce} that $\alpha = 2.5$ and $\beta = 1$ holds. 
We stress, however, that the obtained exponents may still be affected
by rather large numerical errors. In Ref.\ \onlinecite{Kirschner}, a general expression for the multi-particle band-edge singularities is derived for a simple one-dimensional model of hardcore bosons 
with nearest-neighbor hopping. The result reads
\begin{align}
S_n \propto \omega^{\frac{n^2-3}{2}} \quad \mathrm{for} \, \, n>1,
\end{align}, but close to the extrema of the dispersion. 
For the three-particle case, this yields 
$S_n \propto \omega^{3}$ which agrees well with our results for $Q=\pi$ and $Q=\pi/2$, 
but differs for $Q=0$. 
The discrepancy in the latter case may be due to the more complex structure of the
dispersion in the effective Hamiltonian for the TFIM which includes longer range hopping processes.

\subsection{$S^{yy}$ channel}
\label{subsec:dsf_yy}

\begin{figure}
\centering
\includegraphics[width=0.5\textwidth]{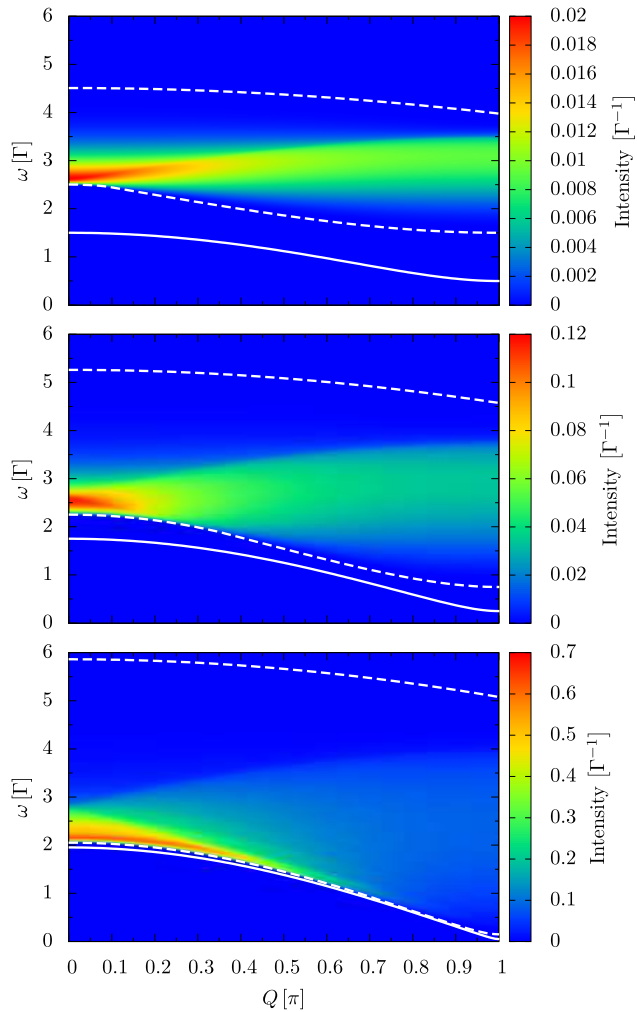}
\caption{(Color online) DSF $S_3^{yy}{(\omega,Q)}$ for the parameters $J=\Gamma$ (top), $J=1.5\Gamma$ (center) and $J=1.9\Gamma$ (bottom). The maximum range for the Lanczos algorithm is $d_\mathrm{max}=100$ sites, the continued fraction was evaluated to a depth of $50$ and then terminated by the square root terminator. The color indicates the spectral density, see legend to the right. The dispersion is indicated by the white solid line. The upper and lower edge of the three-particle continuum are indicated by white dashed lines. All results are computed in order 38.}
\label{image:DSF_yy_3p_j101519}
\end{figure}

As in the  $S^{xx}$ channel, the $S^{yy}$ channel consists of $1,3,5, \dots$ particle contributions.
Overview plots for the three-particle DSF obtained by the CUT in order 38 
are found in Fig.\ \ref{image:DSF_yy_3p_j101519}. In these plots, the three-particle continuum is plotted
in dependence of the total momentum $Q$ and the energy $\omega$. 
For small $J$, the $S^{yy}$ channel looks similar to the $S^{xx}$ channel. The only difference is the absolute 
weight because the three-particle continuum in the $S^{yy}$ channel gains weight sooner{,i.e., for smaller $J/(2\Gamma)$,} than in the $S^{xx}$ channel.

For higher values of $J$, there are already qualitative differences between the $S^{xx}$ and the
$S^{yy}$ channel. Most of the spectral weight is still concentrated at the lower band-edge of the continuum.  
No spectral weight is gained in the region of the critical wave vector $Q=\pi$ which constitutes a major
difference to the $S^{xx}$ channel, see Fig.\ \ref{image:DSF_xx_3p_j101519}. 

\begin{figure}
\centering
\includegraphics[width=0.5\textwidth]{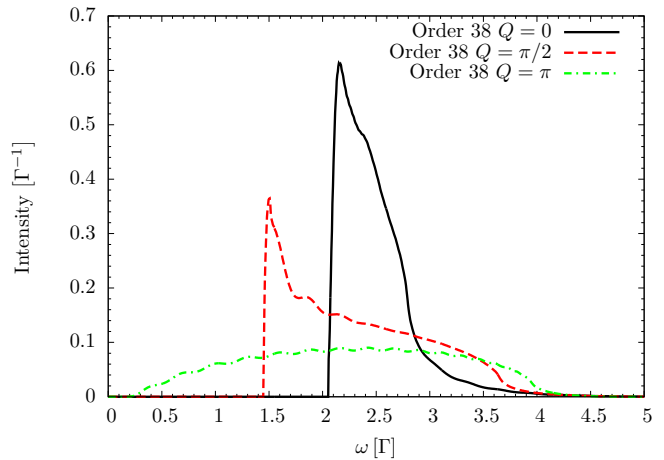}
\caption{(Color online) DSF $S_3^{yy}{(\omega,Q)}$ for the parameter $J=1.9\Gamma$ for three chosen total momenta $Q$. The maximum range for the Lanczos algorithm is $d_\mathrm{max}=300$ sites, the continued fraction was evaluated to a depth of $100$ and then terminated by the square root terminator.}
\label{image:DSF_yy_3p_j19_q_fixed}
\end{figure}

Scans of $S_3^{yy}$ at fixed $Q$ are shown in Fig.\ \ref{image:DSF_yy_3p_j19_q_fixed} for
$J=1.9\Gamma$ and momenta $Q=0$, $Q=\pi/2$ and $Q=\pi$. For $Q \leq \pi/2$,  most of the spectral weight is concentrated at the lower band-edge. This changes distinctly for $Q \geq \pi/2$, especially for $Q \approx \pi$. Here spectral weight is spread rather equally over frequency space. We also observe some wiggling
which is  due to finite order errors.

The values for the band-edge singularities obtained by fits as described before in the
channels $S^{zz}$ and $S_3^{xx}$ are given in Tab.\ \ref{tab.BandEdge_Sing_yy}. 
They mostly equal  those obtained for in the $S^{xx}$ channel within numerical errors. 
Only the case $Q=0$ differs. {We deduce that $\alpha=3$ and $\beta=1$ holds for $S_3^{yy}$ generally}.

\begin{table}
\centering
 \begin{tabular}{|c|c|c|}
\hline
$Q$ & $\alpha$ & $\beta$ \\
\hline
0	   & 2.9 $\pm$ 0.1 & 1.0 $\pm$ 0.1     \\
$\pi/2$ &   2.9 $\pm$ 0.2   & 2.7 $\pm$ 0.2 	  \\
$\pi$  &    2.9 $\pm$ 0.1  & 2.8  $\pm$ 0.2      \\     
\hline
\end{tabular}
\caption{Exponents for the Band Edge Singularities for $S_3^{yy}(\omega,Q)$ for $J=1.5\Gamma$. The errors are determined from the fits using the Levenberg-Marquardt algorithm \cite{Levenberg, Marquardt}.}
\label{tab.BandEdge_Sing_yy}
\end{table}

\section{Conclusion} 
\label{sec:conclusion}

{Summarizing}, we showed that the one-dimensional transverse field 
Ising model (1D TFIM) in the high field limit 
can be expressed in terms of string operators which form an
algebra which is closed under commutation{, which agrees with
the previous finding by Jha and Valatin \cite{jha73}}. 
This property allowed us to solve
the 1D TFIM in the high field limit to very high accuracy by continuous
unitary transformations without resorting to the Jordan-Wigner transformation
to non-interacting fermions. Note that the solution provided formally also 
covers the low field limit due to the duality of the model.

The only remaining restriction in the presented solution is the truncation in
a given order $n$ in the ratio $x=J/(2\Gamma)$ of the Ising coupling $J$ to the field
strength $\Gamma$. But due to the closure of the string algebra the number of
terms to be tracked grows only linearly in the order $n$ so that very high orders up
to $n=256$ can be achieved. Thus, accurate results for all practical purposes could 
be obtained. The order corresponds directly to the range of physical processes which are 
included if the lattice spacing is set to unity. We employed the recently developed
deepCUT approach which is perturbatively correct in the targeted order and 
 provides a robust extrapolation beyond this order \cite{epCUT}.

High orders are accessible for the Hamiltonian and all observables which belong to the
string algebra. They cannot be obtained for observables which do not belong to the string 
algebra such as the longitudinal spin components. For this reason, the longitudinal components 
could be unitarily transformed {only} up to order $38$.

We gauged the results in computing various static properties such
as the ground state energy, the transverse magnetization, and the
one-particle contribution to the equal time structure factors $S_1^{\alpha\alpha}(Q)$
for $\alpha= x$ and $\alpha=y$. 
The first two quantities can be compared directly to the analytically accessible results
via Jordan-Wigner transformation. The 
one-particle contribution to the equal time structure factors 
has been conjectured by Hamer and co-workers by series expansions and
strongly underlined by the mapping to a two-dimensional classical Ising model 
for which the exact results are known \cite{HamerTFIM}. 

Similarly, we computed dynamic quantities such as the one-particle
dispersion and the momentum- and frequency-resolved diagonal dynamic
structure factors. The dispersion and the transverse structure factor
can again be gauged against the analytical result obtained in terms
of non-interacting fermions. The longitudinal structure factors
are much more difficult to address because their excitation
operators are highly non-local in terms of the non-interacting 
fermions. While the one-particle contribution can be derived from 
the static one-particle structure factor and the exactly known
dispersion there are no results for the next important three-particle contributions
for general coupling $x\le 1$.  Only in the scaling region around the quantum
phase transition results for the three-particle contributions {in frequency domain} exist \cite{Vaidya19781}.
Our results are reliable further away from the scaling region so that
they are complementary to the existing information.
{The equation of motion approach pursued by Perk and Au-Yang \cite{perk09} provides
information on the correlations in the time and real space domain. But so far
no analysis with frequency and momentum resolution has been performed.}

The presented theoretical three-particle data for the static and the dynamic
structure factor provides predictions where in 
momentum and frequency space one can expect significant three-particle signal.
This information may guide future experimental searches for many-particle
contributions. Concretely, our results show that 
the $S^{yy}$ channel is considerably better suited for such searches than
the $S^{xx}$ channel. In the $S^{xx}$ channel  the single
particle contributions dominates over the multi-particle contributions
except very close to the quantum phase transition.

Moreover, we found that the spectral weight in the three-particle
dynamic structure factors is concentrated close
to the lower band-edge if the parameters are such 
that the system is not too far away from criticality. 
Further away from criticality the main response is rather featureless 
and hardly displays a dependence on the total momentum $Q$.
Then the spectral weight is still concentrated  close to the lower bandedge around $Q=0$,
while it is spread out in the middle of the band around $Q=\pi$.

Our approach can be pursued further for all one-dimensional models
of which the Hamilton operators can be expressed within the string algebra.
Further investigations for other response functions are possible as well.

\begin{acknowledgments}
We thank Nils Drescher, Mohsen Hafez, Frederik Keim, Holger Krull, and Joachim Stolze  for useful discussions {and Jacques Perk for bringing the series of papers based on his equation of motion
to our attention.}
We acknowledge financial support of the Helmholtz Virtual Institute ``New states of matter and their excitations''.
\end{acknowledgments}

\appendix

\section{Closure of the string algebra}
\label{app:a}

Here we show that the string algebra is closed under commutation. This means that the commutator of two string operators can again be written as a linear combination of string operators.
First, we show that on a chain two string operators commute if neither of their  start-/end-operators 
are on the same site.
Without loss of generality this means
\begin{subequations}
\label{eq:Vanishing_Comms}
\begin{align}
\label{eq:firstcomm}
0=\left[ O_{j,n}^{\phi \epsilon}, O_{l,m}^{\chi \xi} \right]&, \quad l > j, \quad l+m < j+n ,\\
\label{eq:midcomm}
0=\left[ O_{j,n}^{\phi \epsilon}, O_{l,m}^{\chi \xi} \right]&, \quad l > j, \quad l+m > j+n ,\\
\label{eq:lastcomm}
0=\left[ O_{j,n}^{\phi \epsilon}, O_{l,m}^{\chi \xi} \right]&, \quad l > j+n ,
\end{align}
\end{subequations}
for $\phi, \epsilon, \chi, \xi \in \left\lbrace +,- \right\rbrace$. The last commutator \eqref{eq:lastcomm}
is zero because operators acting on completely different sites always commute in a bosonic algebra. A simple calculation yields for the first commutator \eqref{eq:firstcomm}
\begin{subequations}
\begin{align}
\left[ O_{j,n}^{\phi \epsilon}, O_{l,m}^{\chi \xi} \right] &\propto 
\left[ \sigma_l^z \sigma_{l+m}^z, \sigma_l^\chi \sigma_{l+m}^\xi \right] , \\
&= \left[ \sigma_l^z \sigma_{l+m}^z, \sigma_l^\chi \right] \sigma_{l+m}^\xi \\&+ \sigma_l^\chi \left[ \sigma_l^z \sigma_{l+m}^z, \sigma_{l+m}^\xi \right] \nonumber , \\
 &= 2 \left\lbrace (\chi) (\xi) \sigma_l^\chi \sigma_{l+m}^\xi - (\chi) (\xi) \sigma_l^\chi \sigma_{l+m}^\xi \right\rbrace , \\
 &= 0 ,
\end{align}
\end{subequations}
using $\left[ \sigma^z , \sigma^\chi \right] = \chi 2 \sigma^\chi$, $\sigma^z \sigma^\chi = \chi \sigma^\chi$ and $\sigma^\chi \sigma^z = - \chi \sigma^\chi$, if all operators act on the same site.
Analogous calculations yield that \eqref{eq:midcomm} holds as well.

The remaining contributions consist of commutators where either the start- and/or end-operators are on the same site. The start- and/or end-operators on the same site of two string operators need to be different
because otherwise the identities $\sigma^+ \sigma^+ = \sigma^- \sigma^- = 0$ imply
a vanishing result.
Explicit calculations yield for the non-vanishing commutators
\begin{align}
\left[ O_{j,n}^{\phi \epsilon}, O_{j,m}^{-\phi \xi} \right] &= \xi O_{j+m,n-m}^{\xi \epsilon} ,  
\end{align}
with $m<n$ and
\begin{subequations}
\begin{align}
\left[ O_{j,n}^{\phi \epsilon}, O_{j+n,m}^{-\epsilon \xi} \right] &= \epsilon O_{j,n+m}^{\phi \xi}, \\
\left[ O_{j,n}^{\phi \epsilon}, O_{j+n,0} \right] &= - \epsilon 2 O_{j,n}^{\phi \epsilon} , \\
\left[ O_{j,n}^{\phi \epsilon} ,  O_{j,n}^{-\phi -\epsilon} \right] &= \frac{\phi}{2} O_{j,0} + \frac{\epsilon}{2} O_{j+n,0} .
\end{align}
\end{subequations}
Explicit calculations for the translationally invariant string operators yield the following set of commutator relations for the case $n, m \in \mathbb{N}^+, n < m$
\begin{subequations}
\label{eq.T_m1}
\begin{align} 
\left[ T_n^{++}, T_m^{--} \right] &= T_{n+m}^{+-} + T_{n+m}^{-+} - T_{m-n}^{+-} - T_{m-n}^{-+} ,  \\
\left[ T_n^{++}, T_m^{+-} \right] &= T_{n+m}^{++} - T_{m-n}^{++} , \\
\left[ T_n^{++}, T_m^{-+} \right] &= T_{n+m}^{++} - T_{m-n}^{++} , \\
\left[ T_m^{++}, T_n^{+-} \right] &= T_{n+m}^{++} + T_{m-n}^{++} , \\
\left[ T_m^{++}, T_n^{-+} \right] &= T_{n+m}^{++} + T_{m-n}^{++} , \\
\left[ T_n^{++}, T_0^{\phantom{-+}} \right]      &= -4 T_n^{++} , \\
\left[ T_n^{++}, T_m^{++} \right] &= 0 ,
\end{align}
\end{subequations}
and for the case $n=m$
\begin{subequations}
\label{eq.T_m2}
\begin{align}
\left[ T_m^{++}, T_m^{--} \right] &= T_{2m}^{+-} + T_{2m}^{-+} + T_0 ,\\
\left[ T_m^{++}, T_m^{+-} \right] &= T_{2m}^{++} ,\\
\left[ T_m^{++}, T_m^{-+} \right] &= T_{2m}^{++} ,\\
\left[ T_m^{++}, T_m^{++} \right] &= 0,
\end{align}
\end{subequations}
which are all linear combinations of string operators. Contributions with $n>m$ are
 also included by exchange of the arguments in the commutators. 
 This concludes the derivation of the closure of the string algebra.

\section{Proof of infinite order flow equation}\label{app:b}

To prove the expression \eqref{eq:flow_infinite} we proceed in two steps. First, we show that all kinds of string operators of arbitrary range will be created during the flow. Next we show which contributions occur in the DES.

Our starting point for step one is the Hamiltonian of the TFIM in string operators in Eq.\ \eqref{eq.tfim_in_strings}. By induction we show that once we have a complete set of operators of maximum range $n$, $T_0, T_1^{\pm \pm}, T_2^{\pm \pm}, \dots T_n^{\pm \pm}$ we can create a new complete set of operators of range $n+1$ by commutation with a string pair-creation-operator,
\begin{subequations}
\begin{align}
\left[ T_n^{++}, T_1^{--} \right] &= T_{n+1}^{+-} + T_{n+1}^{-+} - T_{n-1}^{+-} - T_{n-1}^{-+} ,\\
\left[ T_n^{++}, T_1^{+-} \right] &= T_{n+1}^{++} - T_{n-1}^{++} ,\\
\left[ T_n^{--}, T_1^{-+} \right] &= -T_{n+1}^{--} + T_{n-1}^{--} .
\end{align}
\end{subequations}

Thereby, we created the string operators of range $n+1$.
Because the Hamiltonian in Eq.\ \eqref{eq.tfim_in_strings} already comprises a complete set of range one we can deduce that all ranges $n \in \mathbb{N}^+$ will be created during the flow.
Hence, we can conclude for the generator of the TFIM
\begin{align}
\eta = \sum\limits_{n=1}^{\infty} t_n^{++} \left(T_n^{++} - T_n^{--} \right) .
\end{align}
For step two we consider the relations in Eq.\ \eqref{eq.T_m1} and Eq.\ \eqref{eq.T_m2}. We start with the contributions to the operator $T_0$. Such contributions are created only in the case $m=n$. 
For a given range $n$ there are two contributions from the commutators 
\begin{subequations}
\begin{align}
\left[ T_n^{++}, T_n^{--} \right] = T_0 + \dots, \\
\left[ T_n^{--}, T_n^{++} \right] = -T_0 + \dots ,
\end{align}
\end{subequations}
both with prefactor one. Note that $T_n^{++}$ and $T_n^{--}$ have the same prefactor up to a sign due to hermiticity/anti-hermiticity. 
Finally, these considerations yield
\begin{align}
\partial_l t_0^{\phantom{+-}} &= 2 \sum\limits_{n=1}^{\infty} \left(t_n^{++}\right)^2
\end{align}
 {for the flow equation for the prefactor of $T_0$.}

Next, we consider the operator $T_m^{+-}$ and $T_m^{-+}$, respectively. They are created by two different kinds of commutators. For $k+l=m$
\begin{subequations}
\begin{align}
\left[ T_k^{++}, T_l^{--} \right] = T_m^{+-} + \dots , \\
\left[ T_k^{--}, T_l^{++} \right] = -T_m^{+-} + \dots ,
\end{align}
\end{subequations}
and for $|k-l|=m$
\begin{subequations}
\begin{align}
\left[ T_k^{++}, T_l^{--} \right] = -T_m^{+-} + \dots  , \\
\left[ T_k^{--}, T_l^{++} \right] = T_m^{+-} + \dots  ,
\end{align}
\end{subequations}
with prefactor one. Note that the operator $T_m^{-+}$ have the same prefactor as $T_m^{+-}$. These calculations yield
\begin{align}
\partial_l t_m^{+-} &= 2 \sum\limits_{k,l}^{k+l=m} t_k^{++} t_l^{++}  - 2 \sum\limits_{k,l}^{|k-l|=m} t_k^{++} t_l^{++} .
\end{align}
Last we consider the operator $T_m^{++}$ and $T_m^{--}$, respectively. They are created by three different kinds of commutators. For $k+l=m$
\begin{subequations}
\begin{align}
\left[ T_k^{++}, T_l^{+-} \right] = T_m^{++} + \dots , \\
\left[ T_k^{++}, T_l^{-+} \right] = T_m^{++} + \dots ,
\end{align}
\end{subequations}
for $|k-l|=m$
\begin{subequations}
\begin{align}
\left[ T_k^{++}, T_l^{+-} \right] = \mathrm{sgn}(k-l) T_m^{++}  + \dots  ,\\
\left[ T_k^{++}, T_l^{-+} \right] = \mathrm{sgn}(k-l) T_m^{++}  + \dots ,
\end{align}
\end{subequations}
where the $\mathrm{sign}$ function stems from the different signs in the cases $\left[ T_n^{++}, T_m^{+-} \right]$ and $\left[ T_m^{++}, T_n^{+-} \right]$ in Eq.\ \eqref{eq.T_m1}. 
Finally, the third case is given by
\begin{align}
\left[ T_m^{++}, T_0 \right] = -4 T_m^{++}.
\end{align} 

Now we can write down the complete flow equation for the prefactor $t_m^{++}$
\begin{subequations}
\begin{align}
\partial_l t_m^{++} &= -4 t_m^{++} t_0 + 2 \sum\limits_{k,l}^{k+l=m} t_k^{++} t_l^{+-}, \nonumber \\ &+ 2 \sum\limits_{k,l}^{|k-l|=m} \mathrm{sgn}(k-l) t_k^{++} t_l^{+-} ,
\end{align}
\end{subequations}
which concludes our derivation for the flow equation for infinite order.

\section{{Calculation of $S_\mathrm{eff}^\alpha(Q)  | 0 \rangle$}}
\label{app:c}

We start from Eq.\ \eqref{eq.DSF_CUT}. To apply the Lanczos algorithm we need to calculate
\begin{align}
S_\mathrm{eff}^\alpha(Q)  | 0 \rangle .
\end{align}
We split the vector into its components of different particle number.
For the two-particle structure factor the state 
\begin{align}
S_\mathrm{eff}\big|_0^2(Q) \ket{0} &= \frac{1}{\sqrt{N}} \sum\limits_{r,d_0,d_1,j} e^{i Q r} \nonumber \\ &\cdot s_{\mathrm{eff},j}^{d_0,d_1} \ket{r+d_0,r+d_0+d_1} ,
\end{align}
with $d_1>0$ must be considered. The sum over $j$ {addresses} all operators that create
an excitation at $r+d_0$ and another at $r+d_0+d_1$ which are different in their content of factors $\sigma^z_i$ at various sites.
The index $j$ is used to distinguish them. 
In contrast, in a strict multi-particle representation there would be only one operator.
Shifting the exponent by $d_0+d_1/2$, the center of mass, results in the expression
\begin{subequations}
\begin{align}
S_\mathrm{eff}\big|_0^2(Q) \ket{0} &= \sum\limits_{d_0,d_1,j}  e^{-i Q (d_0+d_1/2)}  s_{\mathrm{eff},j}^{d_0,d_1} \nonumber \\ &\cdot \underbrace{\frac{1}{\sqrt{N}} \sum\limits_{r} e^{i Q (r+d_0+d_1/2)} \ket{r+d_0,r+d_0+d_1}}_{:= \ket{Q,d_1}} , \\
&= \sum\limits_{d_0,d_1,j}  \underbrace{e^{-i Q (d_0+d_1/2)} s_{\mathrm{eff},j}^{d_0,d_1}}_{:=s_{\mathrm{eff},j}^{d_0,d_1}(Q)} \ket{Q,d_1} ,\\
&= \sum\limits_{d_0,d_1,j} s_{\mathrm{eff},j}^{d_0,d_1}(Q) \ket{Q,d_1} ,
\end{align}
\end{subequations}
where we have introduced $\ket{Q,d_1}$ which is the Fourier transformation of a two-particle state with distance $d_1$.

For the three-particle structure factor the state 
\begin{align}
S_\mathrm{eff}\big|_0^3(Q) \ket{0} &= \frac{1}{\sqrt{N}} \sum\limits_{r,d_0,d_1,d_2,j} e^{i Q r}
s_{\mathrm{eff},j}^{d_0,d_1,d_2} \nonumber \\ &\cdot \ket{r+d_0,r+d_0+d_1,r+d_0+d_1+d_2} ,
\end{align}
with $d_1>0$ and $d_2>0$ must be considered. Shifting the exponent by $d_0+2d_1/3+d_2/3$, the center of mass, results in the expression
\begin{widetext}
\begin{subequations}
\begin{align}
S_\mathrm{eff}\big|_0^3(Q) \ket{0} &= \sum\limits_{d_0,d_1,d_2,j}  e^{-i Q (d_0+2d_1/3+d_2/3)}  s_{\mathrm{eff},j}^{d_0,d_1,d_2} \nonumber \\ &\cdot \underbrace{\frac{1}{\sqrt{N}} \sum\limits_{r} e^{i Q (r+d_0+2d_1/3+d_2/3)} \ket{r+d_0,r+d_0+d_1,r+d_0+d_1+d_2}}_{:= \ket{Q,d_1,d_2}} , \\
&= \sum\limits_{d_0,d_1,d_2,j}  \underbrace{e^{-i Q (d_0+2d_1/3+d_2/3)}  s_{\mathrm{eff},j}^{d_0,d_1,d_2}}_{:=s_{\mathrm{eff},j}^{d_0,d_1,d_2}(Q)} \ket{Q,d_1,d_2} ,\\
&= \sum\limits_{d_0,d_1,d_2,j} s_{\mathrm{eff},j}^{d_0,d_1,d_2}(Q) \ket{Q,d_1,d_2} ,
\end{align}
\end{subequations}
\end{widetext}
where we introduced $\ket{Q,d_1,d_2}$, which is the Fourier transformation of a three-particle state with distance $d_1$ between the first two particles and distance $d_2$ between the second two particles.

\section{Action of the effective Hamiltonian}
\label{app:d}

To apply the Lanczos algorithm we need to know the action of the effective Hamiltonian on the two- and three-particle states, calculated in App.\ \ref{app:c}. We stress that after the CUT there are no terms that violate particle-number conservation. We analyze the action of the operators $T_0, T_n^{+-},T_n^{-+}$ separately. Starting with the simple operator $T_0$ yields
\begin{widetext}
\begin{subequations}
\begin{align}
t_0(\infty) T_0 \ket{Q,d_1} &= t_0(\infty) T_0 \frac{1}{\sqrt{N}} \sum\limits_{r} e^{i Q (r+d_0+d_1/2)} \ket{r+d_0,r+d_0+d_1} , \\
				&= t_0(\infty) (-N+4) \frac{1}{\sqrt{N}} \sum\limits_{r} e^{i Q (r+d_0+d_1/2)} \ket{r+d_0,r+d_0+d_1} , \\
				&=  (E_0 + 4 t_0(\infty)) \ket{Q,d_1} .
\end{align}
\end{subequations}
\end{widetext}
Next, we analyze the action of the operator $T_n^{+-}$
\begin{widetext}
\begin{subequations}
\begin{align}
\sum\limits_{n} t_n^{+-}(\infty) T_n^{+-} \ket{Q,d_1} &= \sum\limits_{n} (-1)^{n-1} t_n^{+-}(\infty) e^{i Q n/2} \ket{Q,d_1+n} \nonumber \\
&+ \sum\limits_{n<d_1} (-1)^{n-1} t_n^{+-}(\infty) e^{i Q n/2} \ket{Q,d_1-n} \\
&+ \sum\limits_{n>d_1} (-1)^{n} t_n^{+-}(\infty) e^{i Q n/2} \ket{Q,n-d_1} , \nonumber
\end{align}
\end{subequations}
\end{widetext}
and of the operator $T_n^{-+}$
\begin{widetext}
\begin{subequations}
\begin{align}
\sum\limits_{n} t_n^{+-}(\infty) T_n^{-+} \ket{Q,d_1} &= \sum\limits_{n} (-1)^{n-1} t_n^{+-}(\infty) e^{-i Q n/2} \ket{Q,d_1+n} \nonumber \\
&+ \sum\limits_{n<d_1} (-1)^{n-1} t_n^{+-}(\infty) e^{-i Q n/2} \ket{Q,d_1-n} \\
&+ \sum\limits_{n>d_1} (-1)^{n} t_n^{+-}(\infty) e^{-i Q n/2} \ket{Q,n-d_1} . \nonumber
\end{align}
\end{subequations}
\end{widetext}
Note the different signs of the second and third term due to the {properties
of the string operator}. 

{Similarly} to the two-particle state we examine the action of the effective Hamiltonian on the three-particle state. The simple operator $T_0$ yields
\begin{subequations}
\begin{align}
t_0(\infty) T_0 \ket{Q,d_1,d_2}	&= t_0(\infty) (-N+6) \ket{Q,d_1,d_2} ,\\
				&=  (E_0 + 6 t_0(\infty)) \ket{Q,d_1,d_2} .
\end{align}
\end{subequations}
Next, we analyze the action of the operator $T_n^{+-}$
\begin{widetext}
\begin{align}
\sum\limits_{n} t_n^{+-}(\infty) T_n^{+-} \ket{Q,d_1,d_2} &= \sum\limits_{n} (-1)^{n-1} t_n^{+-}(\infty) e^{i Q n/3} \ket{Q,d_1+n,d_2} \nonumber \\
&+ \sum\limits_{n<d_1} (-1)^{n-1} t_n^{+-}(\infty) e^{i Q n/3} \ket{Q,d_1-n,d_2+n} \nonumber \\
&+ \sum\limits_{n>d_1} (-1)^{n} t_n^{+-}(\infty) e^{i Q n/3} \ket{Q,n-d_1,d_2+d_1} \\
&+ \sum\limits_{n<d_2} (-1)^{n-1} t_n^{+-}(\infty) e^{i Q n/3} \ket{Q,d_1,d_2-n} \nonumber \\
&+ \sum\limits_{d_1+d_2>n>d_2} (-1)^{n} t_n^{+-}(\infty) e^{i Q n/3} \ket{Q,d_1+d_2-n,n-d_2)} \nonumber \\
&+ \sum\limits_{n>d_1+d_2} (-1)^{n-1} t_n^{+-}(\infty) e^{i Q n/3} \ket{Q,n-d_1-d_2,d_1)} , \nonumber
\end{align}
\end{widetext}
and of the operator $T_n^{-+}$
\begin{widetext}
\begin{align}
\sum\limits_{n} t_n^{+-}(\infty) T_n^{-+} \ket{Q,d_1,d_2} &= \sum\limits_{n} (-1)^{n-1} t_n^{+-}(\infty) e^{-i Q n/3} \ket{Q,d_1,d_2+n}  \nonumber \\
&+ \sum\limits_{n<d_2} (-1)^{n-1} t_n^{+-}(\infty) e^{-i Q n/3} \ket{Q,d_1+n,d_2-n} \nonumber \\
&+ \sum\limits_{n>d_2} (-1)^{n} t_n^{+-}(\infty) e^{-i Q n/3} \ket{Q,d_1+d_2,n-d_2} \\
&+ \sum\limits_{n<d_1} (-1)^{n-1} t_n^{+-}(\infty) e^{-i Q n/3} \ket{Q,d_1-n,d_2} \nonumber  \\
&+ \sum\limits_{d_1+d_2>n>d_1} (-1)^{n} t_n^{+-}(\infty) e^{-i Q n/3} \ket{Q,n-d_1,d_1+d_2-n} \nonumber \\
&+ \sum\limits_{n>d_1+d_2} (-1)^{n-1} t_n^{+-}(\infty) e^{-i Q n/3} \ket{Q,d_2,n-d_1-d_2}. \nonumber
\end{align}
\end{widetext}


\begin{thebibliography}{58}%
\makeatletter
\providecommand \@ifxundefined [1]{%
 \@ifx{#1\undefined}
}%
\providecommand \@ifnum [1]{%
 \ifnum #1\expandafter \@firstoftwo
 \else \expandafter \@secondoftwo
 \fi
}%
\providecommand \@ifx [1]{%
 \ifx #1\expandafter \@firstoftwo
 \else \expandafter \@secondoftwo
 \fi
}%
\providecommand \natexlab [1]{#1}%
\providecommand \enquote  [1]{``#1''}%
\providecommand \bibnamefont  [1]{#1}%
\providecommand \bibfnamefont [1]{#1}%
\providecommand \citenamefont [1]{#1}%
\providecommand \href@noop [0]{\@secondoftwo}%
\providecommand \href [0]{\begingroup \@sanitize@url \@href}%
\providecommand \@href[1]{\@@startlink{#1}\@@href}%
\providecommand \@@href[1]{\endgroup#1\@@endlink}%
\providecommand \@sanitize@url [0]{\catcode `\\12\catcode `\$12\catcode
  `\&12\catcode `\#12\catcode `\^12\catcode `\_12\catcode `\%12\relax}%
\providecommand \@@startlink[1]{}%
\providecommand \@@endlink[0]{}%
\providecommand \url  [0]{\begingroup\@sanitize@url \@url }%
\providecommand \@url [1]{\endgroup\@href {#1}{\urlprefix }}%
\providecommand \urlprefix  [0]{URL }%
\providecommand \Eprint [0]{\href }%
\providecommand \doibase [0]{http://dx.doi.org/}%
\providecommand \selectlanguage [0]{\@gobble}%
\providecommand \bibinfo  [0]{\@secondoftwo}%
\providecommand \bibfield  [0]{\@secondoftwo}%
\providecommand \translation [1]{[#1]}%
\providecommand \BibitemOpen [0]{}%
\providecommand \bibitemStop [0]{}%
\providecommand \bibitemNoStop [0]{.\EOS\space}%
\providecommand \EOS [0]{\spacefactor3000\relax}%
\providecommand \BibitemShut  [1]{\csname bibitem#1\endcsname}%
\let\auto@bib@innerbib\@empty
\bibitem [{\citenamefont {Sachdev}(2001)}]{Sachdev2}%
  \BibitemOpen
  \bibfield  {author} {\bibinfo {author} {\bibfnamefont {S.}~\bibnamefont
  {Sachdev}},\ }\href@noop {} {\emph {\bibinfo {title} {{Quantum Phase
  Transitions}}}}\ (\bibinfo  {publisher} {Cambridge University Press},\
  \bibinfo {year} {2001})\BibitemShut {NoStop}%
\bibitem [{\citenamefont {Katsura}(1962)}]{katsu62}%
  \BibitemOpen
  \bibfield  {author} {\bibinfo {author} {\bibfnamefont {S.}~\bibnamefont
  {Katsura}},\ }\href@noop {} {\bibfield  {journal} {\bibinfo  {journal} {Phys.
  Rev.}\ }\textbf {\bibinfo {volume} {127}},\ \bibinfo {pages} {1508} (\bibinfo
  {year} {1962})}\BibitemShut {NoStop}%
\bibitem [{\citenamefont {Vidal}(2007)}]{vidal07}%
  \BibitemOpen
  \bibfield  {author} {\bibinfo {author} {\bibfnamefont {G.}~\bibnamefont
  {Vidal}},\ }\href@noop {} {\bibfield  {journal} {\bibinfo  {journal} {Phys.
  Rev. Lett.}\ }\textbf {\bibinfo {volume} {98}},\ \bibinfo {pages} {070201}
  (\bibinfo {year} {2007})}\BibitemShut {NoStop}%
\bibitem [{\citenamefont {{H. Y. Yang and K. P. Schmidt}}(2011)}]{gCUT}%
  \BibitemOpen
  \bibfield  {author} {\bibinfo {author} {\bibnamefont {{H. Y. Yang and K. P.
  Schmidt}}},\ }\href@noop {} {\bibfield  {journal} {\bibinfo  {journal}
  {Europhys. Lett.}\ }\textbf {\bibinfo {volume} {94}},\ \bibinfo {pages}
  {17004} (\bibinfo {year} {2011})}\BibitemShut {NoStop}%
\bibitem [{\citenamefont {{P. Jordan and E.
  Wigner}}(1928)}]{JordanWignerTrafo}%
  \BibitemOpen
  \bibfield  {author} {\bibinfo {author} {\bibnamefont {{P. Jordan and E.
  Wigner}}},\ }\href@noop {} {\bibfield  {journal} {\bibinfo  {journal} {Z.
  Phys.}\ }\textbf {\bibinfo {volume} {47}},\ \bibinfo {pages} {631} (\bibinfo
  {year} {1928})}\BibitemShut {NoStop}%
\bibitem [{\citenamefont {Niemeijer}(1967)}]{Niemeijer}%
  \BibitemOpen
  \bibfield  {author} {\bibinfo {author} {\bibfnamefont {T.}~\bibnamefont
  {Niemeijer}},\ }\href@noop {} {\bibfield  {journal} {\bibinfo  {journal}
  {Physica}\ }\textbf {\bibinfo {volume} {36}},\ \bibinfo {pages} {377}
  (\bibinfo {year} {1967})}\BibitemShut {NoStop}%
\bibitem [{\citenamefont {Pfeuty}(1970)}]{Pfeuty}%
  \BibitemOpen
  \bibfield  {author} {\bibinfo {author} {\bibfnamefont {P.}~\bibnamefont
  {Pfeuty}},\ }\href@noop {} {\bibfield  {journal} {\bibinfo  {journal} {Ann.
  of Phys.}\ }\textbf {\bibinfo {volume} {57}},\ \bibinfo {pages} {79}
  (\bibinfo {year} {1970})}\BibitemShut {NoStop}%
\bibitem [{\citenamefont {{O. Derzhko and T. Krokhmalskii}}(1997)}]{DSF2}%
  \BibitemOpen
  \bibfield  {author} {\bibinfo {author} {\bibnamefont {{O. Derzhko and T.
  Krokhmalskii}}},\ }\href {\doibase 10.1103/PhysRevB.56.11659} {\bibfield
  {journal} {\bibinfo  {journal} {Phys. Rev. B}\ }\textbf {\bibinfo {volume}
  {56}},\ \bibinfo {pages} {11659} (\bibinfo {year} {1997})}\BibitemShut
  {NoStop}%
\bibitem [{\citenamefont {{O. Derzhko, T. Verkholyak, T. Krokhmalskii and H.
  B\"uttner}}(2006)}]{DSF}%
  \BibitemOpen
  \bibfield  {author} {\bibinfo {author} {\bibnamefont {{O. Derzhko, T.
  Verkholyak, T. Krokhmalskii and H. B\"uttner}}},\ }\href {\doibase
  10.1103/PhysRevB.73.214407} {\bibfield  {journal} {\bibinfo  {journal} {Phys.
  Rev. B}\ }\textbf {\bibinfo {volume} {73}},\ \bibinfo {pages} {214407}
  (\bibinfo {year} {2006})}\BibitemShut {NoStop}%
\bibitem [{\citenamefont {Perk}(1980)}]{perk80}%
  \BibitemOpen
  \bibfield  {author} {\bibinfo {author} {\bibfnamefont {J.~H.~H.}\
  \bibnamefont {Perk}},\ }\href@noop {} {\bibfield  {journal} {\bibinfo
  {journal} {Phys. Lett. A}\ }\textbf {\bibinfo {volume} {79}},\ \bibinfo
  {pages} {1} (\bibinfo {year} {1980})}\BibitemShut {NoStop}%
\bibitem [{\citenamefont {McCoy}\ \emph
  {et~al.}(1983{\natexlab{a}})\citenamefont {McCoy}, \citenamefont {Perk},\
  and\ \citenamefont {Shrock}}]{mccoy83a}%
  \BibitemOpen
  \bibfield  {author} {\bibinfo {author} {\bibfnamefont {B.~M.}\ \bibnamefont
  {McCoy}}, \bibinfo {author} {\bibfnamefont {J.~H.~H.}\ \bibnamefont {Perk}},
  \ and\ \bibinfo {author} {\bibfnamefont {R.~E.}\ \bibnamefont {Shrock}},\
  }\href@noop {} {\bibfield  {journal} {\bibinfo  {journal} {Nucl. Phys. B}\
  }\textbf {\bibinfo {volume} {220}},\ \bibinfo {pages} {35} (\bibinfo {year}
  {1983}{\natexlab{a}})}\BibitemShut {NoStop}%
\bibitem [{\citenamefont {McCoy}\ \emph
  {et~al.}(1983{\natexlab{b}})\citenamefont {McCoy}, \citenamefont {Perk},\
  and\ \citenamefont {Shrock}}]{mccoy83b}%
  \BibitemOpen
  \bibfield  {author} {\bibinfo {author} {\bibfnamefont {B.~M.}\ \bibnamefont
  {McCoy}}, \bibinfo {author} {\bibfnamefont {J.~H.~H.}\ \bibnamefont {Perk}},
  \ and\ \bibinfo {author} {\bibfnamefont {R.~E.}\ \bibnamefont {Shrock}},\
  }\href@noop {} {\bibfield  {journal} {\bibinfo  {journal} {Nucl. Phys. B}\
  }\textbf {\bibinfo {volume} {220}},\ \bibinfo {pages} {269} (\bibinfo {year}
  {1983}{\natexlab{b}})}\BibitemShut {NoStop}%
\bibitem [{\citenamefont {M\"uller}\ and\ \citenamefont
  {Shrock}(1983)}]{mulle83}%
  \BibitemOpen
  \bibfield  {author} {\bibinfo {author} {\bibfnamefont {G.}~\bibnamefont
  {M\"uller}}\ and\ \bibinfo {author} {\bibfnamefont {R.~E.}\ \bibnamefont
  {Shrock}},\ }\href@noop {} {\bibfield  {journal} {\bibinfo  {journal} {Phys.
  Rev. Lett.}\ }\textbf {\bibinfo {volume} {51}},\ \bibinfo {pages} {219}
  (\bibinfo {year} {1983})}\BibitemShut {NoStop}%
\bibitem [{\citenamefont {M\"uller}\ and\ \citenamefont
  {Shrock}(1984{\natexlab{a}})}]{mulle84a}%
  \BibitemOpen
  \bibfield  {author} {\bibinfo {author} {\bibfnamefont {G.}~\bibnamefont
  {M\"uller}}\ and\ \bibinfo {author} {\bibfnamefont {R.~E.}\ \bibnamefont
  {Shrock}},\ }\href@noop {} {\bibfield  {journal} {\bibinfo  {journal} {Phys.
  Rev. B}\ }\textbf {\bibinfo {volume} {29}},\ \bibinfo {pages} {288} (\bibinfo
  {year} {1984}{\natexlab{a}})}\BibitemShut {NoStop}%
\bibitem [{\citenamefont {M\"uller}\ and\ \citenamefont
  {Shrock}(1984{\natexlab{b}})}]{mulle84b}%
  \BibitemOpen
  \bibfield  {author} {\bibinfo {author} {\bibfnamefont {G.}~\bibnamefont
  {M\"uller}}\ and\ \bibinfo {author} {\bibfnamefont {R.~E.}\ \bibnamefont
  {Shrock}},\ }\href@noop {} {\bibfield  {journal} {\bibinfo  {journal} {Phys.
  Rev. B}\ }\textbf {\bibinfo {volume} {30}},\ \bibinfo {pages} {5254}
  (\bibinfo {year} {1984}{\natexlab{b}})}\BibitemShut {NoStop}%
\bibitem [{\citenamefont {M\"uller}\ and\ \citenamefont
  {Shrock}(1985)}]{mulle85}%
  \BibitemOpen
  \bibfield  {author} {\bibinfo {author} {\bibfnamefont {G.}~\bibnamefont
  {M\"uller}}\ and\ \bibinfo {author} {\bibfnamefont {R.~E.}\ \bibnamefont
  {Shrock}},\ }\href@noop {} {\bibfield  {journal} {\bibinfo  {journal} {Phys.
  Rev. B}\ }\textbf {\bibinfo {volume} {31}},\ \bibinfo {pages} {637} (\bibinfo
  {year} {1985})}\BibitemShut {NoStop}%
\bibitem [{\citenamefont {Perk}\ and\ \citenamefont {Au-Yang}(2009)}]{perk09}%
  \BibitemOpen
  \bibfield  {author} {\bibinfo {author} {\bibfnamefont {J.~H.~H.}\
  \bibnamefont {Perk}}\ and\ \bibinfo {author} {\bibfnamefont {H.}~\bibnamefont
  {Au-Yang}},\ }\href@noop {} {\bibfield  {journal} {\bibinfo  {journal} {J.
  Stat. Phys.}\ }\textbf {\bibinfo {volume} {135}},\ \bibinfo {pages} {599}
  (\bibinfo {year} {2009})}\BibitemShut {NoStop}%
\bibitem [{\citenamefont {Faddeev}\ and\ \citenamefont
  {Takhtajan}(1981)}]{fadde81}%
  \BibitemOpen
  \bibfield  {author} {\bibinfo {author} {\bibfnamefont {L.~D.}\ \bibnamefont
  {Faddeev}}\ and\ \bibinfo {author} {\bibfnamefont {L.~A.}\ \bibnamefont
  {Takhtajan}},\ }\href@noop {} {\bibfield  {journal} {\bibinfo  {journal}
  {Phys. Lett.}\ }\textbf {\bibinfo {volume} {85A}},\ \bibinfo {pages} {375}
  (\bibinfo {year} {1981})}\BibitemShut {NoStop}%
\bibitem [{\citenamefont {Nagler}\ \emph {et~al.}(1991)\citenamefont {Nagler},
  \citenamefont {Tennant}, \citenamefont {Cowley}, \citenamefont {Perring},\
  and\ \citenamefont {Satija}}]{nagle91}%
  \BibitemOpen
  \bibfield  {author} {\bibinfo {author} {\bibfnamefont {S.~E.}\ \bibnamefont
  {Nagler}}, \bibinfo {author} {\bibfnamefont {D.~A.}\ \bibnamefont {Tennant}},
  \bibinfo {author} {\bibfnamefont {R.~A.}\ \bibnamefont {Cowley}}, \bibinfo
  {author} {\bibfnamefont {T.~G.}\ \bibnamefont {Perring}}, \ and\ \bibinfo
  {author} {\bibfnamefont {S.~K.}\ \bibnamefont {Satija}},\ }\href@noop {}
  {\bibfield  {journal} {\bibinfo  {journal} {Phys. Rev. B}\ }\textbf {\bibinfo
  {volume} {44}},\ \bibinfo {pages} {12361} (\bibinfo {year}
  {1991})}\BibitemShut {NoStop}%
\bibitem [{\citenamefont {Dender}\ \emph {et~al.}(1996)\citenamefont {Dender},
  \citenamefont {Davidovi\'c}, \citenamefont {Reich}, \citenamefont {Broholm},
  \citenamefont {Lefmann},\ and\ \citenamefont {Aeppli}}]{dende96}%
  \BibitemOpen
  \bibfield  {author} {\bibinfo {author} {\bibfnamefont {D.~C.}\ \bibnamefont
  {Dender}}, \bibinfo {author} {\bibfnamefont {D.}~\bibnamefont {Davidovi\'c}},
  \bibinfo {author} {\bibfnamefont {D.~H.}\ \bibnamefont {Reich}}, \bibinfo
  {author} {\bibfnamefont {C.}~\bibnamefont {Broholm}}, \bibinfo {author}
  {\bibfnamefont {K.}~\bibnamefont {Lefmann}}, \ and\ \bibinfo {author}
  {\bibfnamefont {G.}~\bibnamefont {Aeppli}},\ }\href@noop {} {\bibfield
  {journal} {\bibinfo  {journal} {Phys. Rev. B}\ }\textbf {\bibinfo {volume}
  {53}},\ \bibinfo {pages} {2583} (\bibinfo {year} {1996})}\BibitemShut
  {NoStop}%
\bibitem [{\citenamefont {Karbach}\ \emph {et~al.}(1997)\citenamefont
  {Karbach}, \citenamefont {M\"uller}, \citenamefont {Bougourzi}, \citenamefont
  {Fledderjohann},\ and\ \citenamefont {M\"utter}}]{karba97}%
  \BibitemOpen
  \bibfield  {author} {\bibinfo {author} {\bibfnamefont {M.}~\bibnamefont
  {Karbach}}, \bibinfo {author} {\bibfnamefont {G.}~\bibnamefont {M\"uller}},
  \bibinfo {author} {\bibfnamefont {A.~H.}\ \bibnamefont {Bougourzi}}, \bibinfo
  {author} {\bibfnamefont {A.}~\bibnamefont {Fledderjohann}}, \ and\ \bibinfo
  {author} {\bibfnamefont {K.~H.}\ \bibnamefont {M\"utter}},\ }\href@noop {}
  {\bibfield  {journal} {\bibinfo  {journal} {Phys. Rev. B}\ }\textbf {\bibinfo
  {volume} {55}},\ \bibinfo {pages} {12510} (\bibinfo {year}
  {1997})}\BibitemShut {NoStop}%
\bibitem [{\citenamefont {Knetter}\ \emph {et~al.}(2001)\citenamefont
  {Knetter}, \citenamefont {Schmidt}, \citenamefont {Gr\"uninger},\ and\
  \citenamefont {Uhrig}}]{knett01b}%
  \BibitemOpen
  \bibfield  {author} {\bibinfo {author} {\bibfnamefont {C.}~\bibnamefont
  {Knetter}}, \bibinfo {author} {\bibfnamefont {K.~P.}\ \bibnamefont
  {Schmidt}}, \bibinfo {author} {\bibfnamefont {M.}~\bibnamefont
  {Gr\"uninger}}, \ and\ \bibinfo {author} {\bibfnamefont {G.~S.}\ \bibnamefont
  {Uhrig}},\ }\href@noop {} {\bibfield  {journal} {\bibinfo  {journal} {Phys.
  Rev. Lett.}\ }\textbf {\bibinfo {volume} {87}},\ \bibinfo {pages} {167204}
  (\bibinfo {year} {2001})}\BibitemShut {NoStop}%
\bibitem [{\citenamefont {Schmidt}\ and\ \citenamefont
  {Uhrig}(2005)}]{schmi05b}%
  \BibitemOpen
  \bibfield  {author} {\bibinfo {author} {\bibfnamefont {K.~P.}\ \bibnamefont
  {Schmidt}}\ and\ \bibinfo {author} {\bibfnamefont {G.~S.}\ \bibnamefont
  {Uhrig}},\ }\href@noop {} {\bibfield  {journal} {\bibinfo  {journal} {Mod.
  Phys. Lett. B}\ }\textbf {\bibinfo {volume} {19}},\ \bibinfo {pages} {1179}
  (\bibinfo {year} {2005})}\BibitemShut {NoStop}%
\bibitem [{\citenamefont {Notbohm}\ \emph {et~al.}(2007)\citenamefont
  {Notbohm}, \citenamefont {Ribeiro}, \citenamefont {Lake}, \citenamefont
  {Tennant}, \citenamefont {Schmidt}, \citenamefont {Uhrig}, \citenamefont
  {Hess}, \citenamefont {Klingeler}, \citenamefont {Behr}, \citenamefont
  {B\"uchner}, \citenamefont {Reehuis}, \citenamefont {Bewley}, \citenamefont
  {Frost}, \citenamefont {Manuel},\ and\ \citenamefont {Eccleston}}]{notbo07}%
  \BibitemOpen
  \bibfield  {author} {\bibinfo {author} {\bibfnamefont {S.}~\bibnamefont
  {Notbohm}}, \bibinfo {author} {\bibfnamefont {P.}~\bibnamefont {Ribeiro}},
  \bibinfo {author} {\bibfnamefont {B.}~\bibnamefont {Lake}}, \bibinfo {author}
  {\bibfnamefont {D.~A.}\ \bibnamefont {Tennant}}, \bibinfo {author}
  {\bibfnamefont {K.~P.}\ \bibnamefont {Schmidt}}, \bibinfo {author}
  {\bibfnamefont {G.~S.}\ \bibnamefont {Uhrig}}, \bibinfo {author}
  {\bibfnamefont {C.}~\bibnamefont {Hess}}, \bibinfo {author} {\bibfnamefont
  {R.}~\bibnamefont {Klingeler}}, \bibinfo {author} {\bibfnamefont
  {G.}~\bibnamefont {Behr}}, \bibinfo {author} {\bibfnamefont {B.}~\bibnamefont
  {B\"uchner}}, \bibinfo {author} {\bibfnamefont {M.}~\bibnamefont {Reehuis}},
  \bibinfo {author} {\bibfnamefont {R.~I.}\ \bibnamefont {Bewley}}, \bibinfo
  {author} {\bibfnamefont {C.~D.}\ \bibnamefont {Frost}}, \bibinfo {author}
  {\bibfnamefont {P.}~\bibnamefont {Manuel}}, \ and\ \bibinfo {author}
  {\bibfnamefont {R.~S.}\ \bibnamefont {Eccleston}},\ }\href@noop {} {\bibfield
   {journal} {\bibinfo  {journal} {Phys. Rev. Lett.}\ }\textbf {\bibinfo
  {volume} {98}},\ \bibinfo {pages} {027403} (\bibinfo {year}
  {2007})}\BibitemShut {NoStop}%
\bibitem [{\citenamefont {Christensen}\ \emph {et~al.}(2007)\citenamefont
  {Christensen}, \citenamefont {R\o{}nnow}, \citenamefont {McMorrow},
  \citenamefont {Harrison}, \citenamefont {Perring}, \citenamefont {Enderle},
  \citenamefont {Coldea}, \citenamefont {Regnault},\ and\ \citenamefont
  {Aeppli}}]{chris07}%
  \BibitemOpen
  \bibfield  {author} {\bibinfo {author} {\bibfnamefont {N.~B.}\ \bibnamefont
  {Christensen}}, \bibinfo {author} {\bibfnamefont {H.~M.}\ \bibnamefont
  {R\o{}nnow}}, \bibinfo {author} {\bibfnamefont {D.~F.}\ \bibnamefont
  {McMorrow}}, \bibinfo {author} {\bibfnamefont {A.}~\bibnamefont {Harrison}},
  \bibinfo {author} {\bibfnamefont {T.~G.}\ \bibnamefont {Perring}}, \bibinfo
  {author} {\bibfnamefont {M.}~\bibnamefont {Enderle}}, \bibinfo {author}
  {\bibfnamefont {R.}~\bibnamefont {Coldea}}, \bibinfo {author} {\bibfnamefont
  {L.~P.}\ \bibnamefont {Regnault}}, \ and\ \bibinfo {author} {\bibfnamefont
  {G.}~\bibnamefont {Aeppli}},\ }\href@noop {} {\bibfield  {journal} {\bibinfo
  {journal} {Proc. Nat. Acad. Sciences}\ }\textbf {\bibinfo {volume} {104}},\
  \bibinfo {pages} {15264} (\bibinfo {year} {2007})}\BibitemShut {NoStop}%
\bibitem [{\citenamefont {Jha}\ and\ \citenamefont {Valatin}(1973)}]{jha73}%
  \BibitemOpen
  \bibfield  {author} {\bibinfo {author} {\bibfnamefont {D.~K.}\ \bibnamefont
  {Jha}}\ and\ \bibinfo {author} {\bibfnamefont {J.~G.}\ \bibnamefont
  {Valatin}},\ }\href@noop {} {\bibfield  {journal} {\bibinfo  {journal} {J.
  Phys. A: Math. Nucl. Gen.}\ }\textbf {\bibinfo {volume} {6}},\ \bibinfo
  {pages} {1679} (\bibinfo {year} {1973})}\BibitemShut {NoStop}%
\bibitem [{\citenamefont {{E. Lieb, T. Schultz and D. Mattis}}(1961)}]{Lieb}%
  \BibitemOpen
  \bibfield  {author} {\bibinfo {author} {\bibnamefont {{E. Lieb, T. Schultz
  and D. Mattis}}},\ }\href@noop {} {\bibfield  {journal} {\bibinfo  {journal}
  {Ann. Phys. (New York)}\ }\textbf {\bibinfo {volume} {16}},\ \bibinfo {pages}
  {407} (\bibinfo {year} {1961})}\BibitemShut {NoStop}%
\bibitem [{\citenamefont {Bogoliubov}(1947)}]{Bogo}%
  \BibitemOpen
  \bibfield  {author} {\bibinfo {author} {\bibfnamefont {N.}~\bibnamefont
  {Bogoliubov}},\ }\href@noop {} {\bibfield  {journal} {\bibinfo  {journal} {J.
  Phys. (USSR)}\ }\textbf {\bibinfo {volume} {11}},\ \bibinfo {pages} {23}
  (\bibinfo {year} {1947})}\BibitemShut {NoStop}%
\bibitem [{\citenamefont {Krones}\ and\ \citenamefont
  {Stolze}(2011)}]{StolzeKrones}%
  \BibitemOpen
  \bibfield  {author} {\bibinfo {author} {\bibfnamefont {J.}~\bibnamefont
  {Krones}}\ and\ \bibinfo {author} {\bibfnamefont {J.}~\bibnamefont
  {Stolze}},\ }\href {\doibase 10.1103/PhysRevB.84.052406} {\bibfield
  {journal} {\bibinfo  {journal} {Phys. Rev. B}\ }\textbf {\bibinfo {volume}
  {84}},\ \bibinfo {pages} {052406} (\bibinfo {year} {2011})}\BibitemShut
  {NoStop}%
\bibitem [{\citenamefont {Lovesey}(1987)}]{Lovesey}%
  \BibitemOpen
  \bibfield  {author} {\bibinfo {author} {\bibfnamefont {S.~W.}\ \bibnamefont
  {Lovesey}},\ }\href@noop {} {\emph {\bibinfo {title} {{Theory of Neutron
  Scattering from Condensed Matter}}}}\ (\bibinfo  {publisher} {Oxford
  University Press},\ \bibinfo {year} {1987})\BibitemShut {NoStop}%
\bibitem [{\citenamefont {{J. H. Taylor and G. M\"uller}}(1985)}]{DSF3}%
  \BibitemOpen
  \bibfield  {author} {\bibinfo {author} {\bibnamefont {{J. H. Taylor and G.
  M\"uller}}},\ }\href@noop {} {\bibfield  {journal} {\bibinfo  {journal}
  {{Physica A}}\ }\textbf {\bibinfo {volume} {130}},\ \bibinfo {pages} {1 }
  (\bibinfo {year} {1985})}\BibitemShut {NoStop}%
\bibitem [{\citenamefont {{C. J. Hamer, J. Oitmaa, Z. Weihong and R.H.
  McKenzie}}(2006)}]{HamerTFIM}%
  \BibitemOpen
  \bibfield  {author} {\bibinfo {author} {\bibnamefont {{C. J. Hamer, J.
  Oitmaa, Z. Weihong and R.H. McKenzie}}},\ }\href {\doibase
  10.1103/PhysRevB.74.060402} {\bibfield  {journal} {\bibinfo  {journal} {Phys.
  Rev. B}\ }\textbf {\bibinfo {volume} {74}},\ \bibinfo {pages} {060402}
  (\bibinfo {year} {2006})}\BibitemShut {NoStop}%
\bibitem [{\citenamefont {Wu}\ \emph {et~al.}(1976)\citenamefont {Wu},
  \citenamefont {McCoy}, \citenamefont {Tracy},\ and\ \citenamefont
  {Barouch}}]{2dIsing1}%
  \BibitemOpen
  \bibfield  {author} {\bibinfo {author} {\bibfnamefont {T.~T.}\ \bibnamefont
  {Wu}}, \bibinfo {author} {\bibfnamefont {B.~M.}\ \bibnamefont {McCoy}},
  \bibinfo {author} {\bibfnamefont {C.~A.}\ \bibnamefont {Tracy}}, \ and\
  \bibinfo {author} {\bibfnamefont {E.}~\bibnamefont {Barouch}},\ }\href
  {\doibase 10.1103/PhysRevB.13.316} {\bibfield  {journal} {\bibinfo  {journal}
  {Phys. Rev. B}\ }\textbf {\bibinfo {volume} {13}},\ \bibinfo {pages} {316}
  (\bibinfo {year} {1976})}\BibitemShut {NoStop}%
\bibitem [{\citenamefont {Orrick}\ \emph {et~al.}(2001)\citenamefont {Orrick},
  \citenamefont {Nickel}, \citenamefont {Guttmann},\ and\ \citenamefont
  {Perk}}]{2dIsing2}%
  \BibitemOpen
  \bibfield  {author} {\bibinfo {author} {\bibfnamefont {W.}~\bibnamefont
  {Orrick}}, \bibinfo {author} {\bibfnamefont {B.}~\bibnamefont {Nickel}},
  \bibinfo {author} {\bibfnamefont {A.}~\bibnamefont {Guttmann}}, \ and\
  \bibinfo {author} {\bibfnamefont {J.}~\bibnamefont {Perk}},\ }\href {\doibase
  10.1023/A:1004850919647} {\bibfield  {journal} {\bibinfo  {journal} {J. Stat.
  Phys.}\ }\textbf {\bibinfo {volume} {102}},\ \bibinfo {pages} {795} (\bibinfo
  {year} {2001})}\BibitemShut {NoStop}%
\bibitem [{\citenamefont {Vaidya}\ and\ \citenamefont
  {Tracy}(1978)}]{Vaidya19781}%
  \BibitemOpen
  \bibfield  {author} {\bibinfo {author} {\bibfnamefont {H.~G.}\ \bibnamefont
  {Vaidya}}\ and\ \bibinfo {author} {\bibfnamefont {C.~A.}\ \bibnamefont
  {Tracy}},\ }\href {\doibase 10.1016/0378-4371(78)90019-5} {\bibfield
  {journal} {\bibinfo  {journal} {Physica A}\ }\textbf {\bibinfo {volume}
  {92}},\ \bibinfo {pages} {1 } (\bibinfo {year} {1978})}\BibitemShut {NoStop}%
\bibitem [{\citenamefont {Wegner}(1994)}]{Wegner}%
  \BibitemOpen
  \bibfield  {author} {\bibinfo {author} {\bibfnamefont {F.}~\bibnamefont
  {Wegner}},\ }\href {\doibase 10.1002/andp.19945060203} {\bibfield  {journal}
  {\bibinfo  {journal} {Ann. Physik}\ }\textbf {\bibinfo {volume} {506}},\
  \bibinfo {pages} {77} (\bibinfo {year} {1994})}\BibitemShut {NoStop}%
\bibitem [{\citenamefont {{S. D. G\l{}azek and K. G.
  Wilson}}(1993)}]{GlazekWilson1}%
  \BibitemOpen
  \bibfield  {author} {\bibinfo {author} {\bibnamefont {{S. D. G\l{}azek and K.
  G. Wilson}}},\ }\href {\doibase 10.1103/PhysRevD.48.5863} {\bibfield
  {journal} {\bibinfo  {journal} {Phys. Rev. D}\ }\textbf {\bibinfo {volume}
  {48}},\ \bibinfo {pages} {5863} (\bibinfo {year} {1993})}\BibitemShut
  {NoStop}%
\bibitem [{\citenamefont {{S. D. G\l{}azek and K. G.
  Wilson}}(1994)}]{GlazekWilson2}%
  \BibitemOpen
  \bibfield  {author} {\bibinfo {author} {\bibnamefont {{S. D. G\l{}azek and K.
  G. Wilson}}},\ }\href {\doibase 10.1103/PhysRevD.49.4214} {\bibfield
  {journal} {\bibinfo  {journal} {Phys. Rev. D}\ }\textbf {\bibinfo {volume}
  {49}},\ \bibinfo {pages} {4214} (\bibinfo {year} {1994})}\BibitemShut
  {NoStop}%
\bibitem [{\citenamefont {Mielke}(1998)}]{Mielke}%
  \BibitemOpen
  \bibfield  {author} {\bibinfo {author} {\bibfnamefont {A.}~\bibnamefont
  {Mielke}},\ }\href@noop {} {\bibfield  {journal} {\bibinfo  {journal} {Eur.
  Phys. J. B}\ }\textbf {\bibinfo {volume} {5}},\ \bibinfo {pages} {605}
  (\bibinfo {year} {1998})}\BibitemShut {NoStop}%
\bibitem [{\citenamefont {{C. Knetter and G.S. Uhrig}}(2000)}]{KnetterUhrig}%
  \BibitemOpen
  \bibfield  {author} {\bibinfo {author} {\bibnamefont {{C. Knetter and G.S.
  Uhrig}}},\ }\href@noop {} {\bibfield  {journal} {\bibinfo  {journal} {Eur.
  Phys. J. B}\ }\textbf {\bibinfo {volume} {13}},\ \bibinfo {pages} {209}
  (\bibinfo {year} {2000})}\BibitemShut {NoStop}%
\bibitem [{\citenamefont {Fischer}\ \emph {et~al.}(2010)\citenamefont
  {Fischer}, \citenamefont {Duffe},\ and\ \citenamefont {Uhrig}}]{fisch10a}%
  \BibitemOpen
  \bibfield  {author} {\bibinfo {author} {\bibfnamefont {T.}~\bibnamefont
  {Fischer}}, \bibinfo {author} {\bibfnamefont {S.}~\bibnamefont {Duffe}}, \
  and\ \bibinfo {author} {\bibfnamefont {G.~S.}\ \bibnamefont {Uhrig}},\
  }\href@noop {} {\bibfield  {journal} {\bibinfo  {journal} {New J. Phys.}\
  }\textbf {\bibinfo {volume} {10}},\ \bibinfo {pages} {033048} (\bibinfo
  {year} {2010})}\BibitemShut {NoStop}%
\bibitem [{\citenamefont {Drescher}\ \emph {et~al.}(2011)\citenamefont
  {Drescher}, \citenamefont {Fischer},\ and\ \citenamefont {Uhrig}}]{dresc11}%
  \BibitemOpen
  \bibfield  {author} {\bibinfo {author} {\bibfnamefont {N.~A.}\ \bibnamefont
  {Drescher}}, \bibinfo {author} {\bibfnamefont {T.}~\bibnamefont {Fischer}}, \
  and\ \bibinfo {author} {\bibfnamefont {G.~S.}\ \bibnamefont {Uhrig}},\
  }\href@noop {} {\bibfield  {journal} {\bibinfo  {journal} {Eur. Phys. J. B}\
  }\textbf {\bibinfo {volume} {79}},\ \bibinfo {pages} {225} (\bibinfo {year}
  {2011})}\BibitemShut {NoStop}%
\bibitem [{\citenamefont {Dusuel}\ and\ \citenamefont
  {Uhrig}(2004)}]{dusue04a}%
  \BibitemOpen
  \bibfield  {author} {\bibinfo {author} {\bibfnamefont {S.}~\bibnamefont
  {Dusuel}}\ and\ \bibinfo {author} {\bibfnamefont {G.~S.}\ \bibnamefont
  {Uhrig}},\ }\href@noop {} {\bibfield  {journal} {\bibinfo  {journal} {J.
  Phys. A: Math. Gen.}\ }\textbf {\bibinfo {volume} {37}},\ \bibinfo {pages}
  {9275} (\bibinfo {year} {2004})}\BibitemShut {NoStop}%
\bibitem [{\citenamefont {Heidbrink}\ and\ \citenamefont
  {Uhrig}(2002)}]{heidb02b}%
  \BibitemOpen
  \bibfield  {author} {\bibinfo {author} {\bibfnamefont {C.~P.}\ \bibnamefont
  {Heidbrink}}\ and\ \bibinfo {author} {\bibfnamefont {G.~S.}\ \bibnamefont
  {Uhrig}},\ }\href@noop {} {\bibfield  {journal} {\bibinfo  {journal} {Eur.
  Phys. J. B}\ }\textbf {\bibinfo {volume} {30}},\ \bibinfo {pages} {443}
  (\bibinfo {year} {2002})}\BibitemShut {NoStop}%
\bibitem [{\citenamefont {{H. Krull, N. A. Drescher and G. S.
  Uhrig}}(2012)}]{epCUT}%
  \BibitemOpen
  \bibfield  {author} {\bibinfo {author} {\bibnamefont {{H. Krull, N. A.
  Drescher and G. S. Uhrig}}},\ }\href {\doibase 10.1103/PhysRevB.86.125113}
  {\bibfield  {journal} {\bibinfo  {journal} {Phys. Rev. B}\ }\textbf {\bibinfo
  {volume} {86}},\ \bibinfo {pages} {125113} (\bibinfo {year}
  {2012})}\BibitemShut {NoStop}%
\bibitem [{\citenamefont {{C. Knetter, K. P. Schmidt and G. S.
  Uhrig}}(2003)}]{StructureOperators}%
  \BibitemOpen
  \bibfield  {author} {\bibinfo {author} {\bibnamefont {{C. Knetter, K. P.
  Schmidt and G. S. Uhrig}}},\ }\href@noop {} {\bibfield  {journal} {\bibinfo
  {journal} {J. Phys. A: Math. Gen.}\ }\textbf {\bibinfo {volume} {36}},\
  \bibinfo {pages} {7889} (\bibinfo {year} {2003})}\BibitemShut {NoStop}%
\bibitem [{\citenamefont {{C. Knetter and G. S.
  Uhrig}}(2004)}]{PhysRevLett.92.027204}%
  \BibitemOpen
  \bibfield  {author} {\bibinfo {author} {\bibnamefont {{C. Knetter and G. S.
  Uhrig}}},\ }\href@noop {} {\bibfield  {journal} {\bibinfo  {journal} {Phys.
  Rev. Lett.}\ }\textbf {\bibinfo {volume} {92}},\ \bibinfo {pages} {027204}
  (\bibinfo {year} {2004})}\BibitemShut {NoStop}%
\bibitem [{\citenamefont {Chandler}(1987)}]{Chandler}%
  \BibitemOpen
  \bibfield  {author} {\bibinfo {author} {\bibfnamefont {D.}~\bibnamefont
  {Chandler}},\ }\href@noop {} {\emph {\bibinfo {title} {{Introduction to
  Modern Statistical Mechanics}}}}\ (\bibinfo  {publisher} {Oxford University
  Press},\ \bibinfo {year} {1987})\BibitemShut {NoStop}%
\bibitem [{\citenamefont {{D. G. Pettifor and D. L. Weaire}}(1985)}]{petti85}%
  \BibitemOpen
  \bibfield  {author} {\bibinfo {author} {\bibnamefont {{D. G. Pettifor and D.
  L. Weaire}}},\ }\href@noop {} {\emph {\bibinfo {title} {{The Recursion Method
  and its Applications}}}},\ Vol.~\bibinfo {volume} {58}\ (\bibinfo
  {publisher} {Springer Verlag},\ \bibinfo {address} {Berlin},\ \bibinfo {year}
  {1985})\BibitemShut {NoStop}%
\bibitem [{\citenamefont {{V. S. Viswanath and G. M\"uller}}(1994)}]{mueller}%
  \BibitemOpen
  \bibfield  {author} {\bibinfo {author} {\bibnamefont {{V. S. Viswanath and G.
  M\"uller}}},\ }\href@noop {} {\emph {\bibinfo {title} {{The Recursion
  Method}}}}\ (\bibinfo  {publisher} {Springer Verlag},\ \bibinfo {address}
  {Berlin},\ \bibinfo {year} {1994})\BibitemShut {NoStop}%
\bibitem [{\citenamefont {{G. S. Uhrig and H. J.
  Schulz}}(1996)}]{PhysRevB.54.R9624}%
  \BibitemOpen
  \bibfield  {author} {\bibinfo {author} {\bibnamefont {{G. S. Uhrig and H. J.
  Schulz}}},\ }\href@noop {} {\bibfield  {journal} {\bibinfo  {journal} {Phys.
  Rev. B}\ }\textbf {\bibinfo {volume} {54}},\ \bibinfo {pages} {R9624}
  (\bibinfo {year} {1996})}\BibitemShut {NoStop}%
\bibitem [{\citenamefont {{G. S. Uhrig and H. J.
  Schulz}}(1998)}]{PhysRevB.58.2900}%
  \BibitemOpen
  \bibfield  {author} {\bibinfo {author} {\bibnamefont {{G. S. Uhrig and H. J.
  Schulz}}},\ }\href@noop {} {\bibfield  {journal} {\bibinfo  {journal} {Phys.
  Rev. B}\ }\textbf {\bibinfo {volume} {58}},\ \bibinfo {pages} {2900}
  (\bibinfo {year} {1998})}\BibitemShut {NoStop}%
\bibitem [{\citenamefont {{K. P. Schmidt, C. Knetter and G.S.
  Uhrig}}(2004)}]{GoetzSpinChain}%
  \BibitemOpen
  \bibfield  {author} {\bibinfo {author} {\bibnamefont {{K. P. Schmidt, C.
  Knetter and G.S. Uhrig}}},\ }\href {\doibase 10.1103/PhysRevB.69.104417}
  {\bibfield  {journal} {\bibinfo  {journal} {Phys. Rev. B}\ }\textbf {\bibinfo
  {volume} {69}},\ \bibinfo {pages} {104417} (\bibinfo {year}
  {2004})}\BibitemShut {NoStop}%
\bibitem [{\citenamefont {{C. Knetter, K. P. Schmidt and G.S.
  Uhrig}}(2003)}]{GoetzSpinLadder}%
  \BibitemOpen
  \bibfield  {author} {\bibinfo {author} {\bibnamefont {{C. Knetter, K. P.
  Schmidt and G.S. Uhrig}}},\ }\href@noop {} {\bibfield  {journal} {\bibinfo
  {journal} {Eur. Phys. J. B}\ }\textbf {\bibinfo {volume} {36}},\ \bibinfo
  {pages} {525} (\bibinfo {year} {2003})}\BibitemShut {NoStop}%
\bibitem [{\citenamefont {{T. Fischer, S. Duffe and G. S.
  Uhrig}}(2011)}]{IPA_CuCl}%
  \BibitemOpen
  \bibfield  {author} {\bibinfo {author} {\bibnamefont {{T. Fischer, S. Duffe
  and G. S. Uhrig}}},\ }\href@noop {} {\bibfield  {journal} {\bibinfo
  {journal} {Europhys. Lett.}\ }\textbf {\bibinfo {volume} {96}},\ \bibinfo
  {pages} {47001} (\bibinfo {year} {2011})}\BibitemShut {NoStop}%
\bibitem [{\citenamefont {Levenberg}(1944)}]{Levenberg}%
  \BibitemOpen
  \bibfield  {author} {\bibinfo {author} {\bibfnamefont {K.}~\bibnamefont
  {Levenberg}},\ }\href@noop {} {\bibfield  {journal} {\bibinfo  {journal}
  {Quarterly of Applied Mathematics}\ }\textbf {\bibinfo {volume} {2}},\
  \bibinfo {pages} {164} (\bibinfo {year} {1944})}\BibitemShut {NoStop}%
\bibitem [{\citenamefont {Marquardt}(1963)}]{Marquardt}%
  \BibitemOpen
  \bibfield  {author} {\bibinfo {author} {\bibfnamefont {D.}~\bibnamefont
  {Marquardt}},\ }\href@noop {} {\bibfield  {journal} {\bibinfo  {journal}
  {SIAM Journal on Applied Mathematics}\ }\textbf {\bibinfo {volume} {11}},\
  \bibinfo {pages} {431} (\bibinfo {year} {1963})}\BibitemShut {NoStop}%
\bibitem [{\citenamefont {Kirschner}(2004)}]{Kirschner}%
  \BibitemOpen
  \bibfield  {author} {\bibinfo {author} {\bibfnamefont {S.}~\bibnamefont
  {Kirschner}},\ }\emph {\bibinfo {title} {{Multi-particle spectral
  densities}}},\ \href@noop {} {\bibinfo {type} {Diplomarbeit, available at
  \url{t1.physik.tu-dortmund.de/uhrig/diploma.html}}},\ \bibinfo  {school}
  {Universit\"at zu K\"oln} (\bibinfo {year} {2004})\BibitemShut {NoStop}%
\end{thebibliography}
%

\end{document}